\documentclass[fleqn,usenatbib]{mnras}

\pdfoutput=1

\usepackage{newtxtext,newtxmath}

\usepackage[T1]{fontenc}
\usepackage{ae,aecompl}
\usepackage{hyperref}

\usepackage{graphicx}	
\usepackage{amsmath}	
\usepackage{amssymb}	

\title[Evolution of boxy/peanut-shaped bulges]{Revealing the cosmic evolution of boxy/peanut-shaped bulges from \textit{HST} COSMOS and SDSS}

\author[S. J. Kruk.]{
Sandor J. Kruk,$^{1,2}${\thanks{E-mail: sandor.kruk@esa.int}\thanks{ESA Research Fellow}}
Peter Erwin,$^{3,4}$
Victor P. Debattista,$^{5}$
Chris Lintott$^{2}$
\\
$^{1}$European Space Agency, ESTEC, Keplerlaan 1, NL-2201 AZ, Noordwijk, The Netherlands\\
$^{2}$Astrophysics Sub-department, University of Oxford, Keble Road, Oxford, OX1 3NP, UK\\
$^{3}$Max-Planck-Insitut f\"ur extraterrestrische Physik, Giessenbachstrasse, D-85748 Garching, Germany\\
$^{4}$Universitats-Sternwarte M\"unchen, Scheinerstrasse 1, D-81679 M\"unchen, Germany\\
$^{5}$Jeremiah Horrocks Institute, University of Central Lancashire, Preston, PR1 2HE, UK
}

\date{Accepted 2019 October 9. Received 2019 October 8; in original form 2019 July 9}

\pubyear{2015}

\begin{document}
\label{firstpage}
\pagerange{\pageref{firstpage}--\pageref{lastpage}}
\maketitle

\begin{abstract}
Vertically thickened bars, observed in the form of boxy/peanut (B/P) bulges, are found in the majority of massive barred disc galaxies in the local Universe, including our own. B/P bulges indicate that their host bars have suffered violent bending instabilities driven by anisotropic velocity distributions. We investigate for the first time how the frequency of B/P bulges in barred galaxies evolves from $z = 1$ to $z
\approx 0$, using a large sample of non-edge-on galaxies with masses $M_{\star} > 10^{10}\:M_{\odot}$, selected from the \textit{HST} COSMOS survey. We find the observed fraction increases from $0^{+3.6}_{-0.0}\%$ at $z = 1$ to $37.8^{+5.4}_{-5.1}$\% at $z = 0.2$. We account for problems identifying B/P bulges in galaxies with low inclinations and unfavourable bar orientations, and due to redshift-dependent observational biases with the help of a sample from the Sloan Digital Sky Survey, matched in resolution, rest frame band, signal-to-noise ratio and stellar mass and analysed in the same fashion. From this, we estimate that the true fraction of barred galaxies with B/P bulges increases from $\sim 10$\% at $z \approx 1$ to $\sim 70\%$ at $z = 0$. In agreement with previous results for nearby galaxies, we find a strong dependence of the presence of a B/P bulge on galaxy stellar mass. This trend is observed in both local and high-redshift galaxies, indicating that it is an important indicator of vertical instabilities across a large fraction of the age of the Universe. We propose that galaxy formation processes regulate the thickness of galaxy discs, which in turn affect which galaxies experience violent bending instabilities of the bar.
\end{abstract}




\section{Introduction}

Many galaxies in the nearby Universe appear vertically thickened in the inner regions when viewed edge-on, appearing as `boxy', `peanut' or even `X'-shaped (e.g. \citealt{Bureau1999,Lutticke2000,Lutticke2000b,Laurikainen2016}). Collectively known as `boxy/peanut bulges' (B/P bulges), these features are the same basic structure, with observed morphology differing due to differences in strength and orientation (\citealt{Combes1981,Combes1990, Athanassoula2002}). They are referred to as bulges following the standard definition of a bulge being a component extending above the plane of the disc. However, the boxy/peanut features are actually part of the bar (see eg. \citealt{Bureau1999}). Understanding the formation and occurrence of boxy/peanut bulges in galaxies has received increased interest since there is now strong evidence that the Milky Way hosts an X-shaped feature, as first reported in observations by the COBE satellite \citep{Dwek1995}, and in many other studies since then (e.g. \citealt{McWilliam2010,Wegg2013,Wegg2015,Ness2016,Shen2016,Clarke2019}). 

Observational studies suggest that bars began forming 8-10 Gyr ago, with the highest observed redshift of a barred galaxy being $z\approx2$ \citep{Simmons2014}. \citet{Sheth2008} and \citet{Melvin2014} have shown that the fraction of disc galaxies observed to have strong bars decreases with redshift, from 35\% at $z=0.2$ to approximately 10\% at $z=1$, suggesting that the majority of bars formed at $z<1$, in agreement with predictions from simulations \citep{Kraljic2012}. At higher redshifts ($z\gtrsim1$), discs tend to be dynamically hotter than in the local Universe, with a higher velocity dispersion compared to ordered rotation \citep{Kassin2012,Girard2018}, which prevents the discs from settling and forming bars \citep{Sellwood2014}. Although numerical modelling of galactic bars predicts the formation of a B/P bulge early in the bar growth stage, 1-2 Gyr after bar formation \citep{Martinez-Valpuesta2004, Martinez-Valpuesta2006, Saha2013}, observationally, the formation time of B/P bulges has not yet been established. 

Several authors have studied the fraction of B/P bulges in nearby edge-on galaxies (e.g. \citealt{Lutticke2000,Yoshino2015,Ciambur2016}), where these structures are most easily identifiable, but until now there have not been any studies of the fraction of B/P bulges in higher redshift galaxies. In the local Universe, \citet{Lutticke2000} found a local fraction of B/P bulges in edge-on galaxies of 45\%, in both optical and infrared, when inspecting a sample of $\sim1,350$ galaxies. In a more recent study using SDSS data, \citet{Yoshino2015} found a lower fraction of 22\%. Although B/P bulges are now clearly understood to be part of the bar, it is difficult to determine a fraction of \textit{barred} galaxies that host B/P structures from studying edge-on galaxies, since determining if an edge-on galaxy is barred or not is difficult. In addition, the orientation of the bar in an edge-on galaxy is important, since a bar viewed end-on will appear to have an elliptical shape, while bars with intermediate orientations and side-on bars will appear to have boxy and peanut shapes, respectively \citep{Lutticke2000}. Therefore, in order to investigate the connection between bars and B/P structures and the evolution from vertically thin to thick structures, studies of face-on or moderately inclined galaxies are needed, where the identification of bars is unambiguous. 

\citet{Erwin2013} showed that it is possible to recognise B/P bulges in moderately inclined galaxies ($i\sim40^{\circ}-70^{\circ}$) by identifying broad, rectangular-shaped isophotes and offset spurs (as shown in their Figure 1), in a sample of 78 local galaxies observed mostly in the infrared. They compared the observations with $N$-body simulations of B/P bulge formed via bar buckling \citep{Debattista2006} to demonstrate that it is possible to identify bars with B/P bulges and without B/P bulges using this method.  

Using this method, and selecting galaxies with well-resolved bars and favourable orientations (moderately inclined with a bar with a position angle $<60^{\circ}$ measured from the galaxy major axis) to detect B/P bulges, \citet{Erwin2017} (hereafter ED17) investigated the fraction of galaxies hosting B/P bulges in 84 local galaxies. They found that B/P bulges are very common in the local Universe, with $\sim50\%$ of barred galaxies hosting B/P bulges, and a strong dependence of the B/P bulge fraction on mass: $\sim80\%$ of galaxies with $M_{\star}>10^{10.4} M_{\odot}$ having B/P bulges compared to only $\sim20\%$ of lower mass galaxies. In addition to moderately inclined galaxies, \citet{Laurikainen2014} suggested that boxy/peanut bulges can also be identified in face-on galaxies, appearing projected as `barlenses' (lens-like structures embedded inside bars; \citealt{Laurikainen2011,Laurikainen2013,Laurikainen2017}). Investigating both inclined and face-on galaxies for the presence of B/P bulges and barlenses, \citet{Li2017} also found a strong dependence on stellar mass in a sample of 264 local disc galaxies from the Carnegie-Irvine Galaxy Survey. It is not yet clear what determines this strong dependence of the presence of a B/P bulge on stellar mass and whether this holds for higher redshift galaxies. 

In this paper, we study the formation time of B/P bulges, assessing whether it is simultaneous or subsequent to bar formation, by investigating how the fraction of B/P bulges evolves with redshift, in a moderately inclined sample of barred galaxies selected from the \textit{Hubble Space Telescope} COSMOS survey and the Sloan Digital Sky Survey (SDSS). Additionally, we study whether the observed strong stellar-mass dependence of B/P bulges in the local Universe is present at higher redshifts. We refer to boxy-, peanut- or X-shaped bulges collectively as \textit{B/P bulges}, and to galaxies without B/P bulges as galaxies with \textit{thin bars}. In Section \ref{data} we describe the selection of a sample of barred galaxies from the \textit{HST} COSMOS survey, a comparison sample from SDSS and artificially redshifted SDSS galaxies, to account for observational biases in identifying B/P bulges. In Section \ref{simulations} we discuss the simulations of the formation of boxy/peanut bulges, and compare the morphological signatures of these features with observations. In Section \ref{classification}, we discuss the method used in visually identifying B/P bulges. In Section \ref{results} we present the results of this project, the B/P bulge fraction with redshift, its stellar-mass dependence and the corrections for observational biases. Finally, in Section \ref{discussion} we discuss the implication of the results in the context of the evolution of barred galaxies. Throughout the paper, we adopt the WMAP Seven-Year Cosmological parameters \citep{Jarosik2011} with ($\Omega_{M},\Omega_{\Lambda},h) = (0.27,0.73,0.71)$

\section{Data}
\label{data}

In this study we use data from the \textit{HST} Cosmic Evolution Survey (COSMOS; \citealt{Koekemoer2007, Scoville2007}) and from SDSS to characterize the redshift evolution of the fraction of barred galaxies hosting B/P bulges. In the following subsections, we describe the observational challenges in detecting B/P bulges, our sample selection from the COSMOS survey and our strategy in matching higher redshift galaxies with local barred galaxies from SDSS.

\subsection{Observational biases}
\label{Observational_biases}

The detection of B/P bulges based on isophotes in higher redshift datasets can be hindered by various effects, particularly: the method used for detecting bars, bandshifting of the rest frame to the observed band, degrading physical resolution of the images and the decrease in signal-to-noise (S/N) with redshift. These will all affect the detected fraction of barred galaxies hosting B/P bulges. To account for these effects, we select a comparison sample of local barred galaxies from SDSS, individually matched to the higher redshift \textit{HST} COSMOS galaxies in each of these observational parameters. We compare these samples to study the redshift evolution of B/P bulges. 

To ensure that similar barred galaxies are selected, or at least that the identification of bars in galaxies is affected by the same biases, we use the same method for selecting barred galaxies in both the COSMOS and SDSS datasets, based on Galaxy Zoo classifications, from the Galaxy Zoo 2 (GZ2; \citealt{Willett2013}) and Galaxy Zoo: Hubble (GZH; \citealt{Willett2017}) projects. In the Galaxy Zoo projects, once the volunteers had classified galaxies as being `featured' or having a disc, and non-edge-on, they were asked whether there was a bar present. The fraction of users positively identifying bars were weighted by their performance on previous images, downweighting a small fraction of classifiers who frequently disagreed with others, thus removing outliers \citep{Willett2013}. We made use of these weighted vote fractions (denoted with $p$) to select a barred sample, as explained in the following subsections.

One observational effect which must be taken into account is the shifting between bands of the galaxies' rest frame emission. It is a known issue that galaxy structure changes as a function of wavelength. Imaging surveys at optical wavelengths only sample the rest frame optical up to $z\sim1$, while at higher redshifts it probes the rest frame ultraviolet, where the images show very different structures \citep{Conselice2004}. This issue is particularly acute for barred galaxies, since bars inhibit star formation and have been shown to have redder colours and thus host an older population of stars than the disc \citep{Kruk2018}. Bars are therefore less evident in the rest frame ultraviolet. Consequently, we limit the redshift of the galaxies selected from COSMOS to $z\leq1$, since above this redshift the observations would probe the rest frame SDSS $u$ band. To account for the changing morphology of galaxies with band, for each of the corresponding SDSS galaxies, we select one of the $g$, $r$ or $i$ band corresponding to the rest frame of the COSMOS galaxies.

The resolution of the images plays an important role in both the detection of bars (e.g. \citealt{Erwin2018}) and identification of B/P bulges. The resolution of the \textit{HST} COSMOS images changes from 0.3 kpc at $z=0.2$ to $\sim0.8$ kpc at $z=1$. Therefore, to account for the detectability of B/P due to degrading resolution, the sample of barred SDSS galaxies was selected to have the same spatial resolution, based on the FWHM of the observations, in the $g$, $r$, or $i$ band, corresponding to the rest frame of the \textit{HST} COSMOS matched galaxy. Additionally, to account for the degrading S/N in the \textit{HST} COSMOS images, we added noise to the SDSS images, by artificially redshifting the SDSS galaxies to the redshift of the corresponding COSMOS galaxy, using the \textsc{ferengi} code \citep{Barden2012}, as detailed in Section \ref{artificial_redshifting}.

Finally, since there is good evidence from the local Universe that the fraction of B/P bulges depends strongly on stellar mass (ED17), we are interested in comparing equivalent local and higher redshift galaxies, which is achieved by matching the COSMOS and SDSS galaxies in stellar mass.

\begin{figure}
\centering
 \includegraphics[width=\columnwidth]{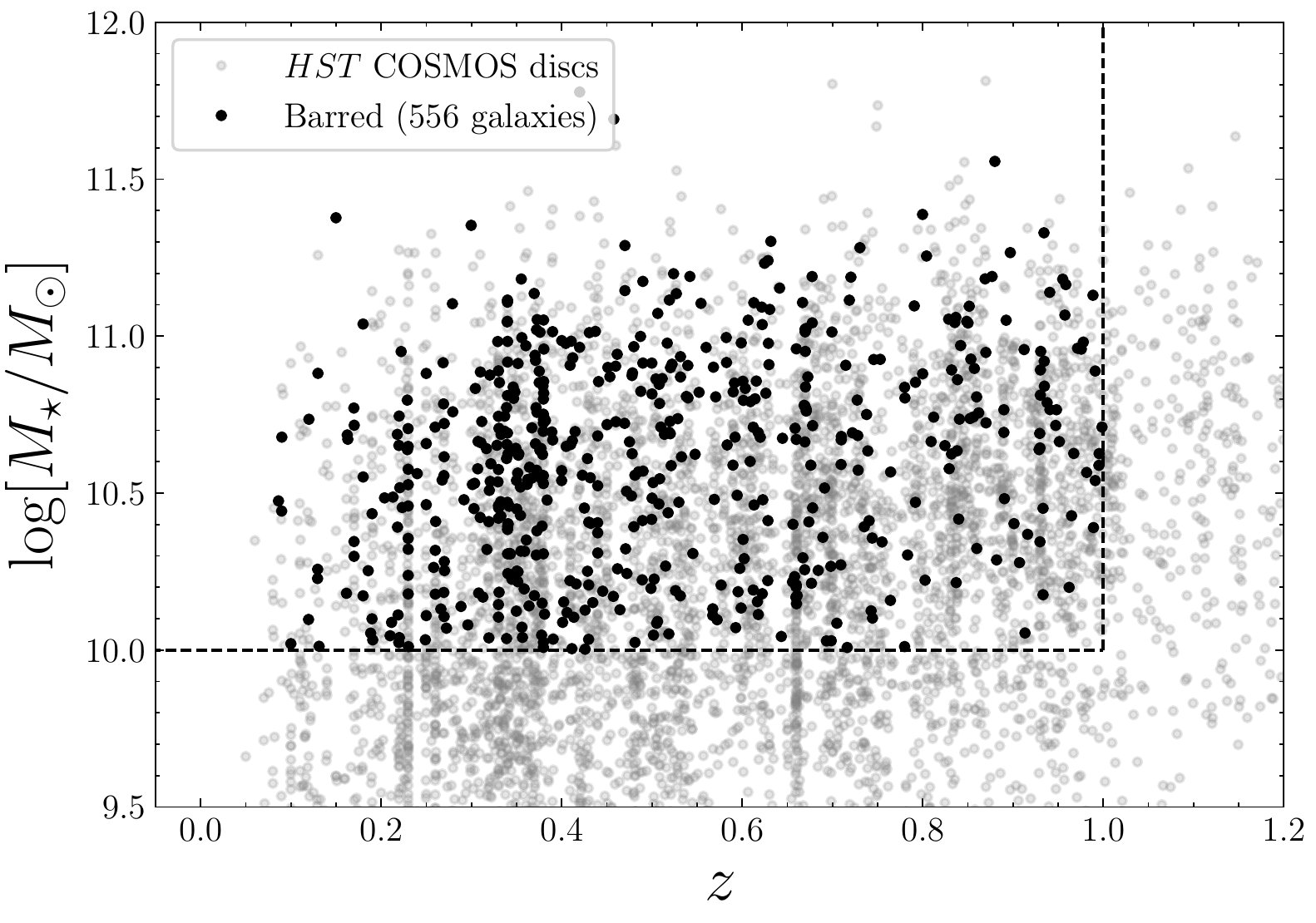}
 \caption[Mass-redshift plot for the selected \textit{HST} galaxies in this study]{Mass-redshift relation for the selected \textit{HST} COSMOS galaxies in this study. The gray points show the disc galaxies identified in \textit{HST} COSMOS and the black points show the moderately inclined barred galaxies with $p_{\mathrm{bar}}>0.5$ as identified in the Galaxy Zoo: Hubble project. Stellar masses and photometric redshifts are from the COSMOS2015 catalogue \citep{Laigle2016}.}
 \label{HST_mass_redshift}
\end{figure}

\begin{figure*}
\centering
\includegraphics[width=1\textwidth]{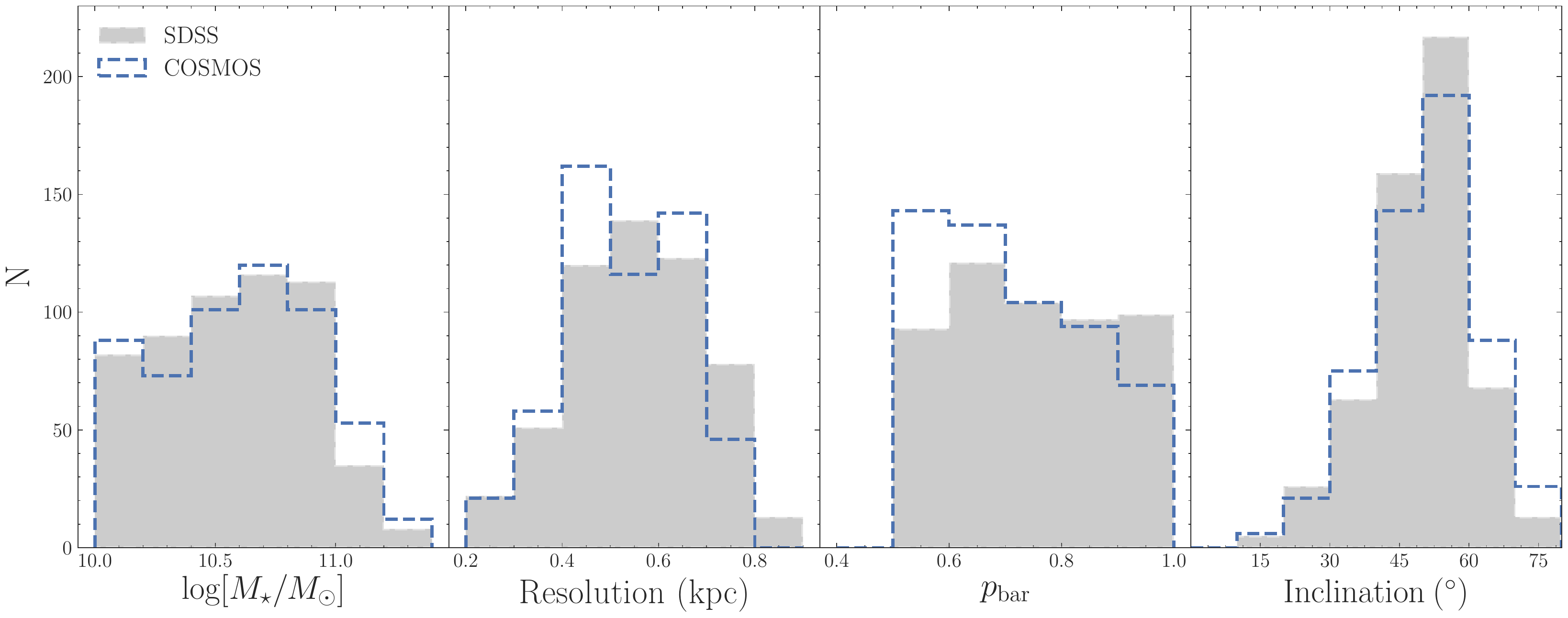}
\caption{Histograms showing the matching in stellar mass, physical resolution and bar likelihood ($p_{\mathrm{bar}}$) for the \textit{HST} COSMOS and SDSS samples. A comparison of the inclinations of the galaxies in the two datasets is shown in the 4${^\mathrm{th}}$ panel. Even though the two samples were not matched in inclination, their distributions are similar.}
\label{HST_SDSS_mass_resolution_pbar}
\end{figure*}

\subsection{\textit{Hubble Space Telescope} - COSMOS data}
\label{COSMOS_data}

The COSMOS survey, covering an area of 1.8 deg$^{2}$ centred at $\alpha=10^{h}00^{m}$ and $\delta=+02^{\circ}12'$, is the largest wide-field survey ever undertaken by \textit{HST}; it was designed to probe the evolution of galaxies, star formation, active galactic nuclei and dark matter with large-scale structure up to $z\sim3$ \citep{Scoville2007}. The \textit{HST} images used in this study were taken with the Advanced Camera for Surveys (ACS) using the $I$-band F814W filter during 590 orbits consisting of 590 pointings with an exposure time of 2028 s each \citep{Koekemoer2007}. The ACS has an excellent spatial resolution of 0.05 arcsec pixel$^{-1}$, and the PSF FWHM of the observations is 0.09$^{\prime \prime}$, corresponding to $\sim0.8$ kpc at $z=1$. 

The COSMOS2015 catalogue \citep{Laigle2016} provides stellar masses and precise photometric redshifts to over half a million objects in the COSMOS field. We use their stellar mass measurements, which are based on BC03 \citep{Bruzual2003} synthetic spectrum fitting (best-fitting template, minimum $\chi^2$, \textit{mass best} value in their catalogue) based on multiwavelength $YJHK_s$ imaging and assuming a Chabrier IMF \citep{Chabrier2003}. The photometric redshifts were obtained by fitting spectral templates to the galaxy SEDs. The values used in this work are the median of the redshift likelihood distribution. The precision of the photometric redshifts of galaxies at $z<1$ is high  $\sigma_{\Delta z/(1+z)}=0.007$ (dispersion measurement; \citealt{Hoaglin1983}), when compared to the measured spectroscopic redshifts of a subsample of galaxies (zCOSMOS; \citealt{Lilly2007}) and the fraction of catastrophic errors ($|z_{\mathrm{phot}}-z_{\mathrm{spec}}|/(1+z_{\mathrm{spec}})>0.15$) is $\eta=0.5\%$ \citep{Laigle2016}.

To select a high-redshift sample of barred galaxies we make use of volunteers' classifications from the Galaxy Zoo citizen science project. 85,000 images of galaxies from COSMOS were classified in the third incarnation of the Galaxy Zoo project, Galaxy Zoo: Hubble  \citep{Willett2017}. Although the COSMOS survey provides only $I$-band F814W images, GZH pseudo-colour images were created based on Subaru telescope $B_J$, $r_{+}$ and $i_{+}$ filters (see \citealt{Griffith2012}). The images have a median of 48 classifications, which were converted to a vote fraction, i.e. the fraction of volunteers that answered positively, for each question in the classification tree (shown in Figure 4 of \citealt{Willett2017}). Following the recommendations of \citet{Willett2017}, we selected galaxies with $p_{\mathrm{features}}>0.23$, $p_{\mathrm{edge-on,no}}>0.25$, $p_{\mathrm{clumpy,no}}>0.30$, resulting in a non-edge-on, non-clumpy disc galaxy sample. These thresholds were found by \citet{Willett2017} to generate samples of galaxies with a purity $>80\%$, based on their expert visual inspection of subsamples. The selection of $p_{\mathrm{edge-on,no}}>0.25$ removes edge-on galaxies from the sample. The $p_{\mathrm{clumpy,no}}>0.30$ threshold in GZH guarantees the exclusion of galaxies that have mostly a clumpy appearance, frequently observed at high redshifts \citep{Elmegreen2007} and for which the bar identification is difficult. We then selected galaxies where at least 20 volunteers answered the bar question, to ensure that the question is well sampled, and where 50\% of them classified the galaxy as being barred ($p_{\mathrm{bar}}>0.5$). This threshold has been shown to select mainly galaxies with strong bars \citep{Masters2011,Skibba2012,Kruk2018}, and, at the resolution of \textit{HST}, with lengths $\geq$ 1-2 kpc. Lower thresholds would increase the sample size, but at the expense of introducing galaxies with uncertain classification to the sample. 

In addition, we selected only galaxies with stellar masses $M_{\star}>10^{10}\:M_{\odot}$ which were in the redshift range $0\leq z\leq1$. This mass threshold is motivated by the findings of ED17 that galaxies with stellar masses $M_{\star}<10^{10} M_{\odot}$ are observed to have a very low fraction of B/P bulges ($f_{\mathrm{B/P}}\lesssim0.1$). The upper redshift limit of $z\approx1$ is set by the F814W band shifting to the rest frame $u$-band, as discussed in the previous subsection, where bar detection has proven to be difficult due to clumpy star formation and lower S/N \citep{Sheth2008,Melvin2014}. This yielded a total of 556 barred COSMOS galaxies. These are shown in the mass-redshift diagram of Figure \ref{HST_mass_redshift}, along with the parent sample of disc galaxies from COSMOS (including unbarred and edge-on systems).

The ACS F814W \textsc{fits} images used for the classification of boxy/peanut bulges are available from the STSci Hubble Legacy Archive\footnote{\href{https://hla.stsci.edu/hlaview.html/ }{https://hla.stsci.edu/hlaview.html/}}.

\subsection{SDSS comparison data}

\begin{figure*}
\centering
 \includegraphics[width=\textwidth]{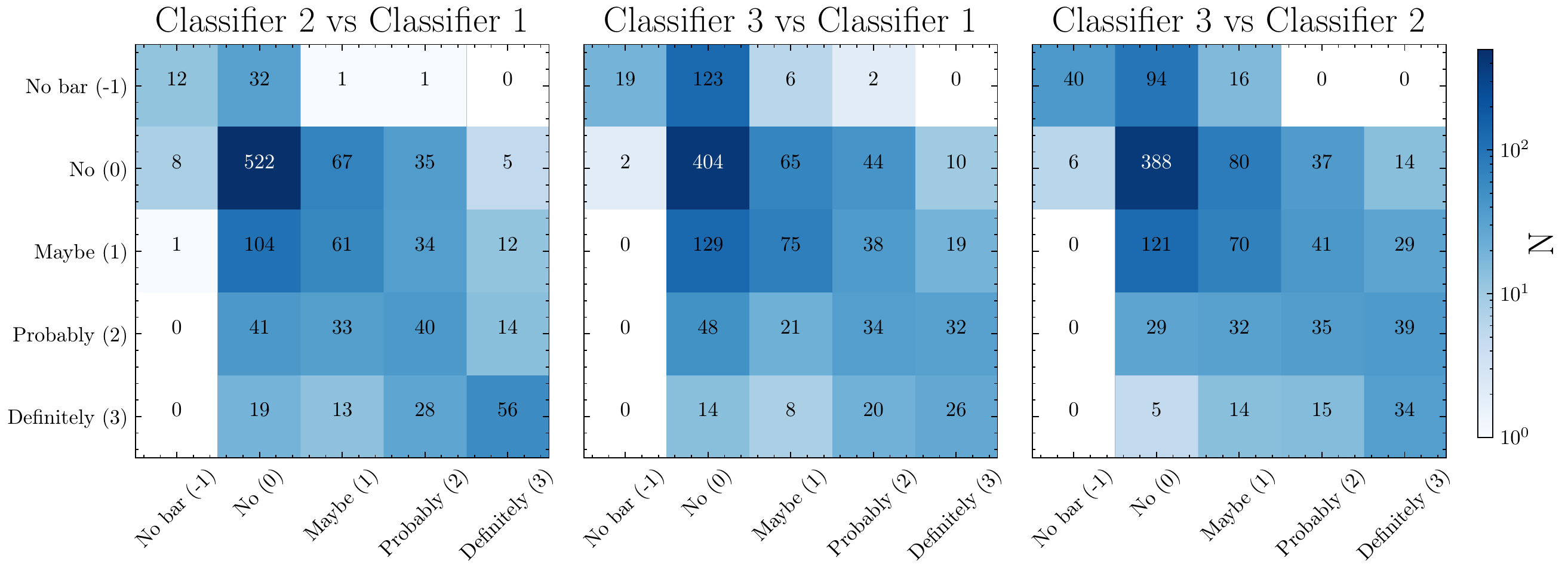}
 \caption[Agreement between the expert classifiers in classifying B/P bulges]{Agreement between the three classifiers for the 1,112 classified COSMOS and SDSS galaxies. The 2D histogram is colour coded, on a logarithmic scale, by the number of galaxies in each bin, with the number displayed in each bin.}
 \label{bp_classifications}
\end{figure*}

\begin{figure}
\centering
 \includegraphics[width=\columnwidth]{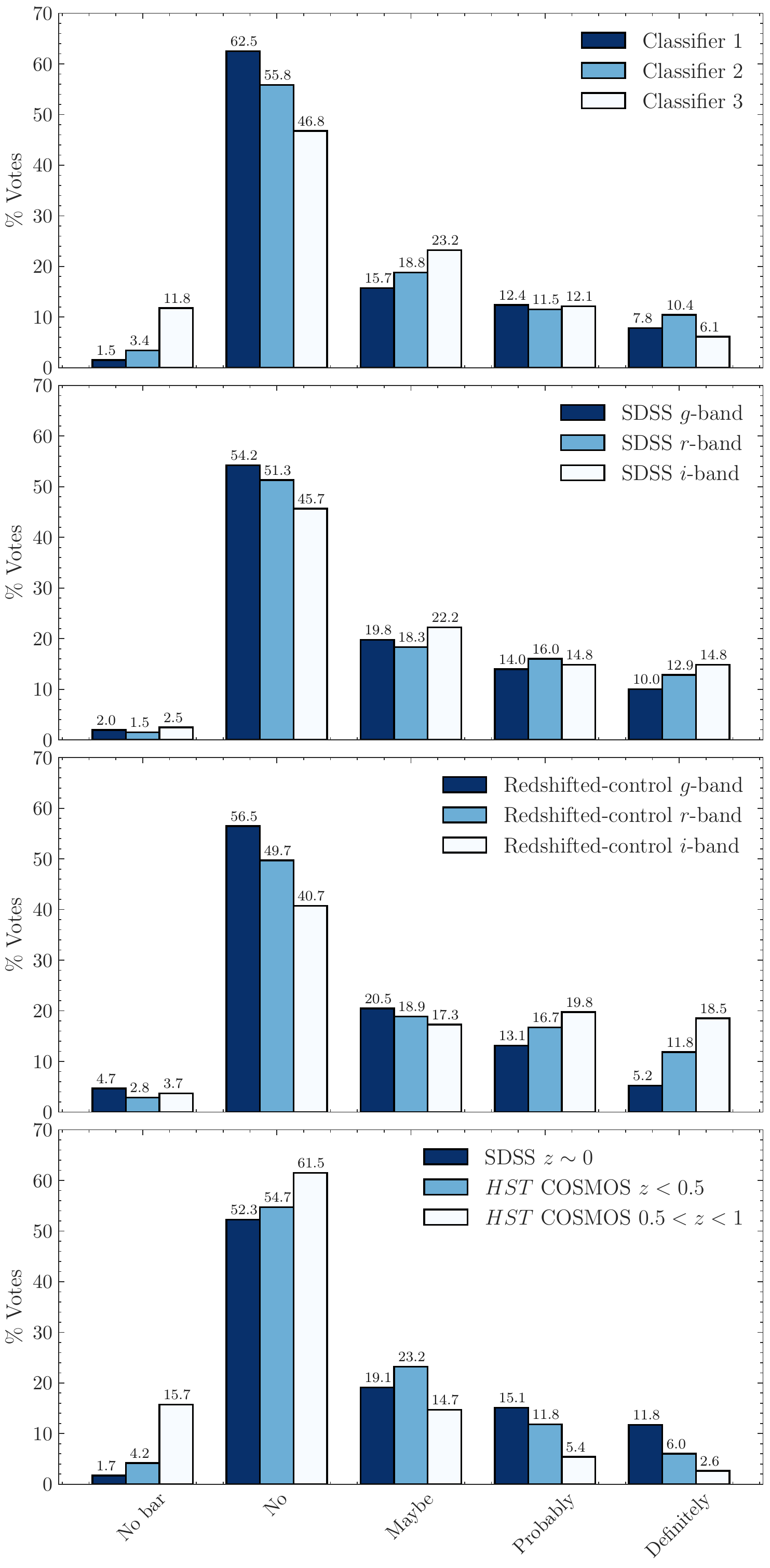}
 \caption{Normalized histograms of the classifications into `No bar', `No B/P', `Maybe B/P',  `Probably B/P' and `Definitely B/P' split by: \textit{top panel} - classifier, \textit{middle panels} - $gri$ band for the $z\approx0$ SDSS data and for the $z=0-1$ \textsc{redshifted}-control data and \textit{bottom panel} - dataset, split by redshift (local SDSS, low-redshift $HST$ COSMOS and high-redshift $HST$ COSMOS). The numbers on the top of the histograms are \% of the total votes in each panel.}
 \label{histogram_classifications}
\end{figure}

In order to study the true redshift evolution of the B/P bulge fraction, from the local Universe to $z\sim1$, we selected a local ($z \approx 0$) sample matched in rest frame band and resolution from SDSS Data Release 10 \citep{Gunn1998,York2000, Ahn2014}, with morphological classifications from the Galaxy Zoo 2 project \citep{Willett2013}. From the initial sample of 243,000 GZ2 galaxies with available spectroscopic redshifts, we selected barred galaxies using similar criteria as for the higher redshift dataset, with the exception of the clumpy question which did not appear in GZ2. Thus, we selected barred galaxies with $p_{\mathrm{features}}>0.23$, $p_{\mathrm{edge-on,no}}>0.25$, $N_{\mathrm{bar}}\geq10$ and $p_{\mathrm{bar}}>0.5$, giving a total of 14,104 galaxies. The lower threshold for the number of classifications for the bar question in SDSS is to account for the slightly lower total number of classifications collected in the GZ2 project compared to GZH. With this selection, the median number of classifications for the bar question is the same for the two samples of barred galaxies (COSMOS and SDSS).  

The median resolution of the SDSS images of the selected galaxies is 1.2$^{\prime\prime}$, 1.13$^{\prime\prime}$ and 1.06$^{\prime\prime}$ in the $g$, $r$, $i$ bands, respectively.  This translates to a physical resolution at the redshift of the SDSS galaxies ranging from 0.1 kpc to $\sim$5 kpc, the upper end of which is worse than the resolution of \textit{HST} at $z\sim1$ (0.8 kpc). To ensure resolutions similar to those obtained with \textit{HST}, we restrict the SDSS sample to galaxies situated within $z\lesssim0.04$ (2,003 barred galaxies). 

Similarly to the COSMOS dataset, only galaxies with $M_{\star}>10^{10} M_{\odot}$ were chosen, where the stellar masses for the SDSS galaxies were taken from the MPA-JHU catalogue \citep{Kauffmann2003a}. From this superset of SDSS barred galaxies, we selected a random sample of 556 unique (without replacement) SDSS galaxies such that each COSMOS galaxy was paired with an SDSS galaxy having stellar mass matching within $\pm 0.2$ dex, $p_{\rm bar}$ matching within $\pm 0.2$, and physical resolution (in the SDSS band corresponding to the rest frame band of the COSMOS image) matching within $\pm 0.2$ kpc in FWHM. This resulted in a final sample containing 556 high redshift galaxies and a control sample of 556 $z\approx0$ galaxies, as shown in Figure \ref{HST_SDSS_mass_resolution_pbar}. The two distributions do not match exactly because of: (1) binning \textendash $ $ individual galaxies in the two samples are matched in the parameters described above, rather than matched in bins; and (2) there are not enough high-mass SDSS barred galaxies with comparable resolution to \textit{HST}, as the more massive galaxies in SDSS are preferentially situated at higher redshifts (due to the larger volume probed). A better match between the two stellar mass distributions could be achieved by sampling with replacement, however this would be at the expense of introducing further discrepancies in the distribution of physical resolutions, since the highest mass SDSS galaxies have the worst resolution images. Nevertheless, we have investigated matching the COSMOS mass distribution by sampling with replacement and all the results are similar to those presented in this paper. Thus, in the analysis we used the SDSS comparison sample matched without replacement, since all the 556 SDSS galaxies were inspected and classified.

Axis ratios for the SDSS galaxies can be estimated from $r$-band exponential model fits in SDSS \citet{Stoughton2002} and from the S\'ersic fits of \citet{Griffith2012} for COSMOS galaxies. The distribution of galaxy inclinations, calculated from the apparent axis ratios ($\cos{i}\approx b/a$), are shown in Figure \ref{HST_SDSS_mass_resolution_pbar} ($4^{th}$ panel). Both the higher redshift COSMOS and local SDSS galaxies in this study are selected to have moderate inclinations (95\% of galaxies have inclinations $i\sim30^\circ-75^\circ$), but both samples also include a small fraction ($\sim$ $5\%$) of face-on galaxies, with $i\lesssim30^{\circ}$. The two distributions are similar, even though matching the two distributions was not a selection criterion.

\begin{figure*}
\centering
\makebox[\textwidth][c]{\includegraphics[width=1\textwidth]{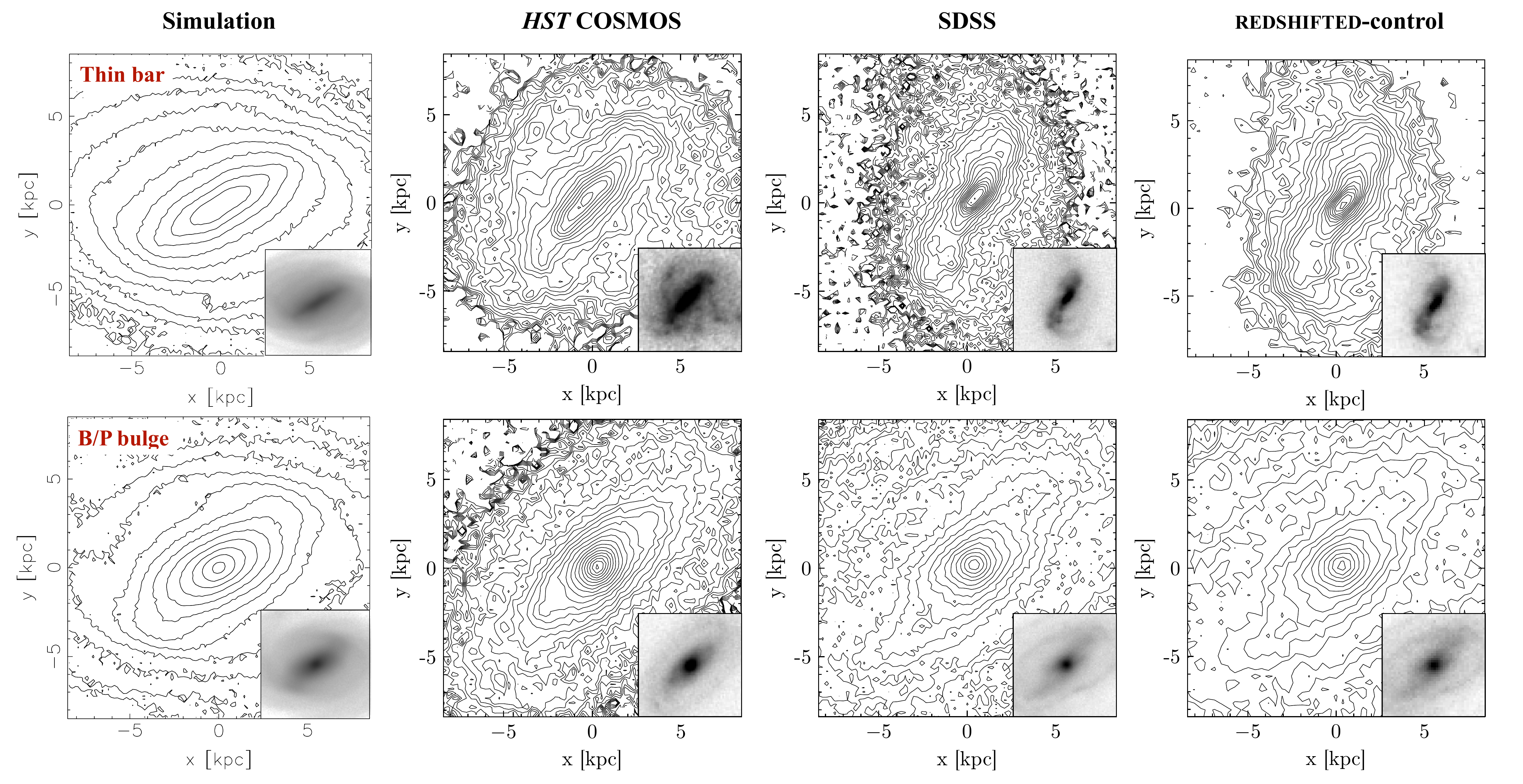}}
\caption{\textit{Top row}: Example of a thin bar (No detected B/P bulge) from $N$-body simulations \citep{Debattista2017}, \textit{HST} COSMOS, SDSS and \textsc{redshifted}-control SDSS data. \textit{Bottom row}: Example of a B/P bulge from $N$-body simulations \citep{Debattista2017} and B/P bulges as identified in COSMOS and SDSS data, following the typical isophotes - boxy isophotes in the centre and offset spurs on either side of the rectangle. The subplots at the bottom-right show the original images of the galaxies from which the contour plots were derived, with an arcsinh stretch applied to the flux.}
\label{examples_simulations}
\end{figure*}

\subsection{Artificially redshifting SDSS galaxies}
\label{artificial_redshifting}

Although the two imaging datasets (SDSS and \textit{HST} COSMOS) were selected to have similar physical resolution, they can differ significantly in S/N ratio. This is for a number of reasons, including different exposure times; however, the most important effect is probably surface brightness dimming at higher redshifts, which means that increasingly higher redshift COSMOS images have increasingly lower S/N relative to both lower redshift COSMOS images and to SDSS images. Since this study concerns galaxies at different redshifts, we need to account for the possible effect of degrading S/N on the detectability of B/P bulges.

We first estimated the S/N per pixel by measuring the flux between 0.1 and 0.6 of the galaxies' effective radii, equivalent, on average, to the region between the bulge and the end of the bar, as found by \citealt{Kruk2018}, in the SDSS and \textit{HST} images. We assumed that the noise is dominated by the source since galaxies are bright and the background is low, this assumption being valid even at the highest redshifts ($z\sim1$). Adding the background to the S/N measurements does not change the measurements significantly. The S/N ratio for the COSMOS galaxies drops below the median SDSS S/N for  $z>0.6$. Therefore, one needs to account for this difference in S/N at the highest redshifts when classifying B/P bulges

To ensure that the S/N ratio is matched in the two datasets, in particular at highest redshifts, we artificially redshifted the SDSS galaxies, using the \textsc{ferengi} code \citep{Barden2008}, to the equivalent redshift of the corresponding COSMOS galaxies. First, we rescale the SDSS images to 0.05 arcsec pixel$^{-1}$ to match the pixel-scale of the COSMOS images, while conserving the physical size of the galaxies. Then, we apply a flux correction, corresponding to the cosmological bolometric surface brightness dimming of $(1+z)^{-4}$. The flux in the images is scaled to the 2028 s COSMOS integration time and Poisson noise (with $\sigma^{2}$ corresponding to the galaxy flux per pixel) was added.  Finally, we place the galaxy on an empty sky region of COSMOS and reclassified all the 556 artificially redshifted SDSS galaxies. Hereafter we refer this sample as the \textsc{redshifted}-control dataset. 

Our artificial redshifting approach is similar to that of other authors (e.g. \citealt{Paulino-Afonso2017,Willett2017}), except that the SDSS images were not convolved with the \textit{HST} ACS PSF because the SDSS data was a priori selected to have the same physical resolution as \textit{HST}. Furthermore, the selection of the SDSS $gri$ band equivalent to the rest frame of ACS F814W already accounts for the potential B/P classification bias due to bandshifting.

\subsection{Simulations}
\label{simulations}

In order to remind the reader of the morphology of a B/P bulge in images of inclined galaxies, we present a simulation which forms such a bulge.  Since \citet{Erwin2013} and ED17 have already presented examples of simulations with these properties, here we present a simulation which they did not consider, in order to enlarge the sample of published models. We consider the model CL described in \citet{Debattista2017}. This is a disc$+$dark matter halo model in which the disc is comprised of `populations' with the same exponential disc scale length but different total mass and, more importantly, different in-plane random motion.  Full details of the model are presented in \citet{Debattista2017} and are not repeated here. Figure 3 of that paper shows the evolution of each of the stellar populations.  \citet{Debattista2017} showed that, while the different populations start out with the same disc scale height, the formation of the bar and its subsequent buckling causes the different discs to thicken by different amounts, with the populations which were initially hotter in-plane ending up more vertically extended.  They referred to this process as kinematic fractionation.  As a simulation with multiple populations in the disc, it is ideal for comparing with real galaxies.  In Figure \ref{examples_simulations} we present images of the simulation after the bar has formed but before buckling (at $t=1$ Gyr) and after it buckles and thickens (at $t=5$ Gyr), forming a B/P-shaped bulge. The boxy centre$+$offset spur morphology is readily apparent at $t=5$ Gyr.  This morphology is absent at $t=1$ Gyr, when the model exhibits only co-aligned thin contours instead.

\section{Classification method}
\label{classification}

The visual identification of B/P bulges we use is based on isophote plots and on the method outlined in \citet{Erwin2013}. Three levels of zoom are used for the inspection, one that encompasses the entire galaxy, an intermediate zoom level and one that in most cases encloses only the bar region.  Square images with side lengths of 30, 60 and 150 pixels (corresponding to 12$^{\prime\prime}$, 24$^{\prime\prime}$ and 1$^{\prime}$ for SDSS and 1.5$^{\prime\prime}$, 3$^{\prime\prime}$ and 7.5$^{\prime\prime}$ for COSMOS, respectively) are used, with a logarithmic scaling of the contour levels. The isophote plots and the three levels of zoom allow us to reliably examine the presence of a B/P bulge in the differently sized galaxies, which is not possible to achieve by inspecting standard RGB colour images. 

At moderate inclinations, the B/P bulge projects to form thicker, often box-shaped, isophotes, and the outer part of the bar projects to form thin offset isophotes referred to as spurs \citep{Erwin2013}, as illustrated in galaxies NGC 1808, NGC 1617, NGC 3992 and simulations A and E in Figure 3 in ED17. The same figure shows the projection effects of the inclination $i$ and difference in position angle between the bar and disc ($\Delta\mathrm{PA_{bar}}$) on the isophotes. If the galaxy does not have a B/P bulge, the projected isophotes are approximately elliptical and the spurs are not distinguishable, as in the case of IC 676. This is also illustrated in Figure \ref{examples_simulations}, where the top row shows a \textit{thin bar} in the simulations, $HST$ COSMOS and SDSS data, while the second row shows a \textit{B/P bulge} (identified based on rectangular isophotes and offset spurs) in the same datasets. 

B/P bulges can also be identified in galaxies with low inclinations, as barlenses \citep{Laurikainen2014, Athanassoula2015, Laurikainen2017, Salo2017}, which appear as lens-like structures with rounder isophotes, typically covering half of the length of the bar. Although our sample contains galaxies which are nearly face-on, we did not attempt to identify B/P bulges on the basis of the presence of barlenses\footnote{Nevertheless, our B/P bulge sample might contain barlenses. Out of 36 galaxies classified as having B/P bulges in \citet{Erwin2017}, 8 (22\%) have been classified as having barlenses in \citet{Buta2015}. Thus, considering that we classified B/P bulges in galaxies based on the same boxy-shaped isophotes and offset spur morphology analysis as in \citet{Erwin2017}, we estimate that our B/P bulge sample contains $\sim20\%$ barlenses.}, as their identification might be affected by the misclassification of classical bulges or large nuclear discs inside bars \citep{Erwin2017}.

\begin{figure*}
\centering
\makebox[\textwidth][c]{\includegraphics[width=1\textwidth]{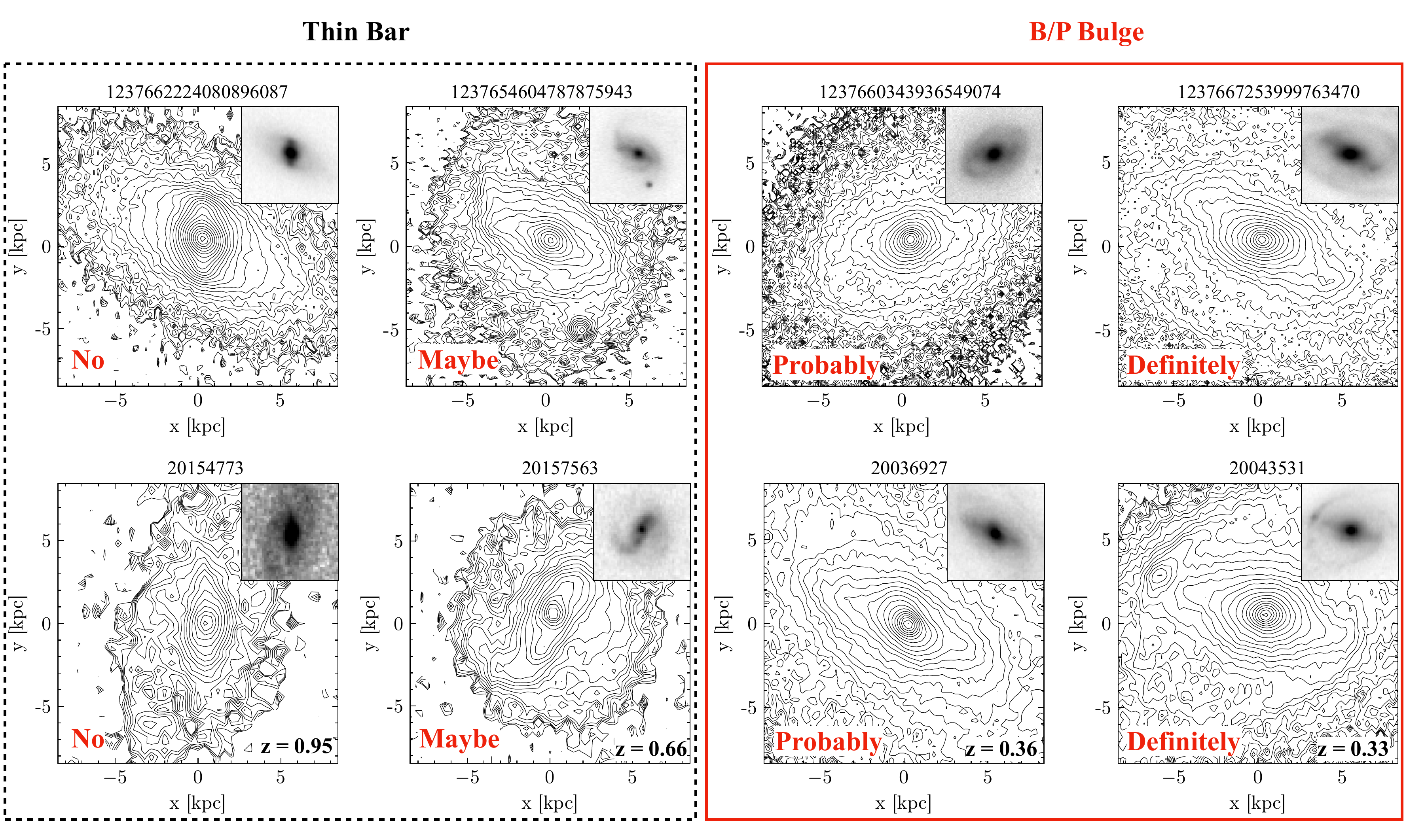}}
\caption[Examples of SDSS and COSMOS galaxies classified as having `No', `Maybe', `Probably' and `Definitely' B/P bulges]{Examples of SDSS (\textit{top row}) and COSMOS (\textit{bottom row}) galaxies classified by at least two out of three classifiers as having  `No', `Maybe', `Probably' and `Definitely' B/P bulges. The B/P bulges were identified based on the boxy isophotes and the offset spurs on either side of the rectangle. Galaxies with a majority of votes for `No' and `Maybe' were classified as having \textit{thin bars} (\textit{left panels}), while galaxies classified with the majority of votes for `Probably' and `Definitely' as \textit{B/P bulges} (\textit{right panels}). The \textit{top-left} SDSS galaxy probably has a B/P bulge, but because the bar is close to the disc minor axis, the isophotes do not show typical B/P bulge signatures, hence it was classified as having a thin bar. The subplots at the top-right show the original images of the galaxies from which the contour plots were derived, with an arcsinh stretch applied to the flux.}
\label{mosaic_bp}
\end{figure*}

\subsection{Zooniverse classification project}

To classify B/P bulges more efficiently we set-up a project using the Zooniverse Panoptes Project Builder\footnote{\href{https://www.zooniverse.org/projects/sandorkruk/b-p-bulge-finder}{https://www.zooniverse.org/projects/sandorkruk/b-p-bulge-finder}}. Instead of a \textit{Yes/No} answer for the question \textit{`Does the galaxy have a B/P bulge?'}, we decided on five possible answers: `Definitely', `Probably', `Maybe', `No' and `No bar' to capture our confidence in classifying the galaxy isophotes. The `No bar' option was added in the case of galaxies without bars that were misclassified in GZ2 or GZH. Each answer was assigned a score: -1 (`No bar'), 0 (`No B/P'), 1 (`Maybe B/P'),  2 (`Probably B/P'), and 3 (`Definitely B/P').

The 556 higher-$z$ galaxies from \textit{HST} (COSMOS sample) and 556 local galaxies from SDSS (SDSS control), along with the 556 redshifted SDSS galaxies (\textsc{redshifted}-control sample) to the \textit{HST} redshifts were classified by three authors $-$ SK (Classifier 1), PE (Classifier 2) and VD (Classifier 3). Apart from answering the \textit{`Does the galaxy have a B/P bulge?'} question, galaxies with buckling bars were identified based on trapezoidal isophotes and spurs along the base of the trapezium (similarly to \citealt{Erwin2016}). These galaxies were classified as \textit{not} having B/P bulges and are the subject of a subsequent paper. Figure \ref{bp_classifications} shows the agreement between the classifiers for the five categories for the COSMOS and SDSS datasets (1,112 galaxies). The majority of the classifications ($\sim400-500$) were for the `No B/P' category. In general, there is a good agreement between the three classifiers for most answers, shown by the darker colours along the diagonal. Most of the scatter is explained by a one step difference in classifications, for example one classifier selecting `Maybe' (or `Probably') and another `No' or `Probably' (or `Definitely'). For 128 galaxies (11\% of the sample), the authors disagree on the classifications, i.e. each chose a different answer for the B/P question. 

We investigated the distributions of the five possible answers in Figure \ref{histogram_classifications}. First, the top panel suggests that the fraction of votes for the five categories for the COSMOS and SDSS datasets is similar for the three classifiers. The second panel shows the distribution of votes for the $z\approx0$ SDSS sample, split between the three $gri$ bands. There is no significant discrepancy between the vote fractions in the three bands, suggesting that as long as rest frame optical images are inspected, the band should not have a significant effect on the identification of B/P bulges. The third panel shows the distribution of votes for the \textsc{redshifted}-control dataset, split into the three bands, which can be compared to the real SDSS images in the second panel. \textsc{redshifted}-control galaxies classified in the $g$ band received fewer `Probably' and `Definitely' votes compared to the $r$ and $i$ bands, and fewer than for the real SDSS dataset in the $g$ band. The $g$ band images were redshifted to higher redshifts compared to the $i$ band, with higher noise, thus the less certain classifications are probably due to degrading S/N in the images. 

Finally, the bottom panel shows the same distributions, split by redshift into local $z\approx0$ SDSS, low-redshift $HST$ COSMOS ($z<0.5$) and high-redshift $HST$ COSMOS ($0.5<z<1$). In this case, the distribution of votes changes between the three categories - the vote fractions for `Probably' and `Definitely' decrease with redshift bin (from 15.1\% to 5.4\% and 11.7\% to 2.6\%, respectively), while for `No' and `No bar' the vote fractions increase (from 52.3\% to 61.7\% and 1.7\% to 15.7\%, respectively). We discuss this further in Section \ref{results}, where we show that this is mostly due to the physical evolution of B/P bulges with redshift. 

\subsection{Galaxies with boxy/peanut bulges}

The scores from each of the three classifications are added together ($\Sigma_{\textrm{score}}$) for each galaxy and the galaxies are split into three categories, depending on the total score: galaxies \textit{without bars} ($\Sigma_{\textrm{score}}<-1$), galaxies with \textit{thin bars} (without detected B/P bulges) ($-1\leq\Sigma_{ \textrm{score}} \leq3$) and barred galaxies with \textit{B/P bulges} ($\Sigma_{\textrm{score}} >3$). Unbarred galaxies are removed from the sample, and the B/P bulge classifications are treated as a binary choice; this allows us to compute the fraction of barred galaxies hosting B/P bulges and to model the observed trends using logistic regression. The boundary between galaxies with B/P bulges and with thin bars ($\Sigma_{\textrm{score}}=3$) corresponds to all three classifications for a galaxy being `Maybe' (which was considered to be a negative answer in this paper). Galaxies with two classifications of `Maybe' and one of `Probably' or any combinations with `Probably' and `Definitely' are considered to have B/P bulges. Examples of galaxies with the majority of votes ($\geq2/3$) for `No', `Maybe', `Probably'  or `Definitely' are shown in Figure \ref{mosaic_bp}. 

We remind the reader that our detection methodology is based on projection effects which are strongest when the galaxy is moderately inclined and when the bar is not too close to the galaxy's minor axis.
This means that some B/P bulges may fail to be detected, with their host galaxies being (erroneously) classified as having thin bars, purely because of orientation effects. This might be the case with the SDSS galaxy in the top-left image of Figure \ref{mosaic_bp}. The sample of thin bars thus contains galaxies with truly thin bars, and galaxies with unfavourable disc and bar relative orientations to identify the B/P bulge based on the isophotes. The B/P fractions determined in Section \ref{bp_fraction_redshift} are therefore \textit{lower limits} on the true B/P fractions. With our methodology, it would be possible to measure the true fraction of B/P bulges only in a sample of galaxies with favourable bar orientations and inclinations. Nevertheless, we attempt to correct for this effect in Section \ref{corrected_fractions}.

Out of the 1,112 classified galaxies (556 COSMOS and 556 SDSS) there are 259 classified as having B/P bulges (89 COSMOS and 170 SDSS), 816 without B/P bulges (436 COSMOS and 380 SDSS) and 37 having no bars (31 COSMOS and 6 SDSS). In the analysis that follows, the 37 galaxies classified as not barred were removed, leaving a total of 525 COSMOS galaxies and 550 SDSS galaxies. Additionally, since in the analysis it is important to have an equal number of COSMOS and SDSS galaxies in order to consistently compare the two datasets, we removed an extra 6 COSMOS galaxies whose corresponding (same mass, $p_{\mathrm{bar}}$ and resolution) SDSS galaxy was classified as having `No bar', and similarly, 30 SDSS galaxies whose corresponding COSMOS galaxies were without bars, leaving a total of 520 COSMOS and 520 SDSS galaxies. For comparison, in the \textsc{redshifted}-control dataset of 556 galaxies there are 152 galaxies classified as having B/P bulges, 395 as not having B/P bulges and 9 as not being barred. The detected fraction of B/P bulges in the COSMOS dataset is thus 17\%, while in SDSS it is 31\% and in the \textsc{redshifted}-control dataset 28\%. 

We investigate the effect of choosing a threshold ($\Sigma_{\textrm{score}}=3$) in separating B/P bulges from thin bars on the results by computing the `likelihood' of having a B/P bulge for each galaxy (by aggregating the different classification options), instead of a binary choice for the B/P bulges. In Appendix \ref{maybe_yes}, we study the redshift evolution of this B/P `likelihood' and show that the results are qualitatively the same as presented in the next section.

\section{Results}
\label{results}

In this section, we investigate the fraction of B/P bulges as a function of redshift and stellar mass, based on our classifications. We refer to the measured fractions of barred galaxies with B/P bulges as the `detected fractions', $f_{\mathrm{B/P,\:det}}$, since the true fractions of B/P bulges are affected by observational biases such as the resolution and S/N of the images, as discussed in Section \ref{Observational_biases}, as well as by the relative orientation of the bar and the disc and the inclination of the galaxies. We account for redshift-dependent observational biases
(band-shifting, degrading resolution and S/N) by using the resolution-, S/N- and stellar-mass--matched \textsc{redshifted}-control dataset. In Section \ref{corrected_fractions}, we also attempt to correct for redshift-independent biases due to galaxy orientations (missing B/P bulges due to low inclinations or unfavourable bar orientations) and compute `corrected fractions', $f_{\mathrm{B/P,\:cor}}$, representing the true fraction of B/P bulges. We present the fractions binned in redshift, in bins of $\Delta z=0.2$. The error bars in the following plots are the 68\% ($1\sigma$) confidence limits from the \citealt{Wilson1927} binomial confidence interval.

\subsection{Boxy/peanut bulge fraction with redshift}
\label{bp_fraction_redshift}

\begin{figure}
\centering
 \includegraphics[width=\columnwidth]{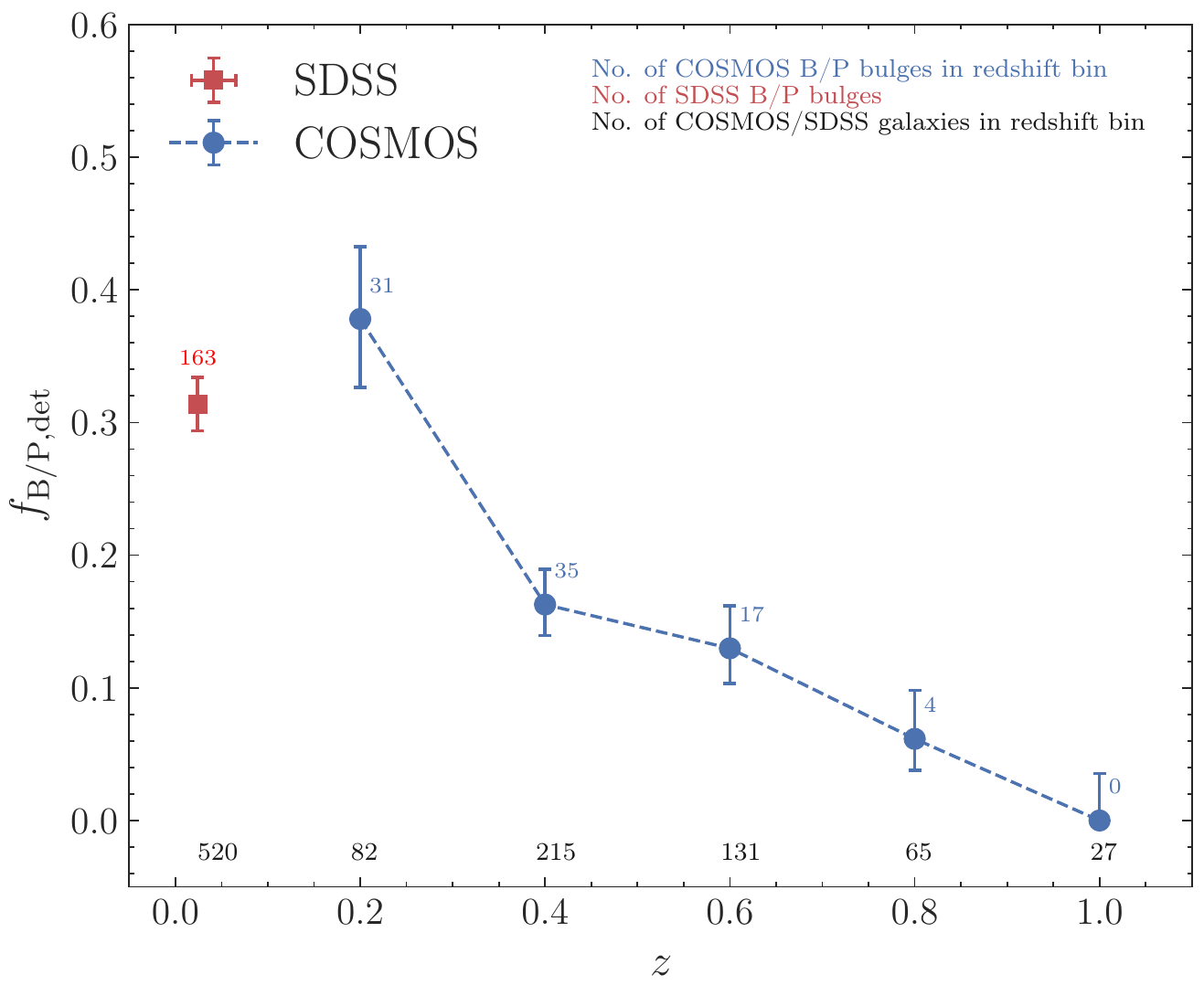}
 \caption{The detected B/P bulge fraction ($f_{\mathrm{B/P,\:det}}$) against redshift for the SDSS and COSMOS barred galaxies, as measured by the three classifiers. The blue points represent the mean B/P bulge fraction of COSMOS galaxies in each redshift bin. The red square shows the local B/P bulge fraction as measured for the SDSS galaxies. The numbers in blue (red) represent the number of B/P bulges in each bin for the COSMOS (SDSS) galaxies, while the numbers at the bottom of the figure represent the total number of galaxies in each bin.}
 \label{BP_vs_redshift}
\end{figure}

\begin{figure}
\centering
 \includegraphics[width=\columnwidth]{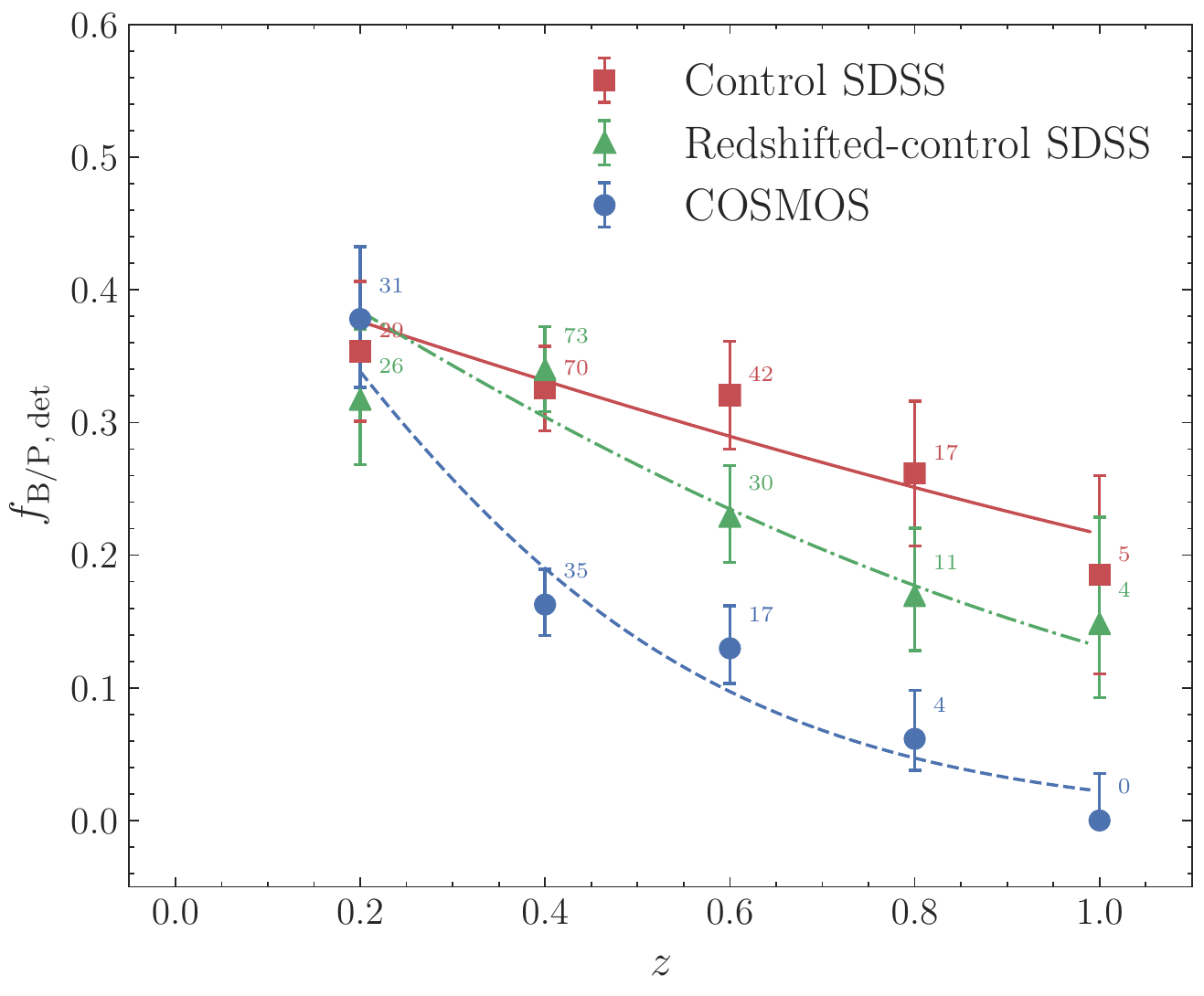}
 \caption{The detected fraction of barred galaxies with B/P bulges for the COSMOS galaxies, as well as for the SDSS and the \textsc{redshifted}-control sample, as a function of the equivalent redshift of the corresponding COSMOS galaxy. The dashed, solid and dash-dotted lines show the logistic fits, obtained using Eq. \ref{eqn}, to the individual galaxies in the three datasets. The decline in the SDSS and the \textsc{redshifted}-control B/P fractions imply that the determined fractions are affected by degrading resolution and degrading S/N. The numbers represent the number of B/P bulges in each bin.}
 \label{BP_fraction_comparison}
\end{figure}

The detected fraction of barred galaxies having boxy/peanut bulges $f_{\mathrm{B/P,\:det}}$ is plotted as a function of redshift in Figure \ref{BP_vs_redshift}. This plot shows some interesting trends. At $z \approx 0$, we have $f_{\mathrm{B/P,\:\mathrm{det,\:SDSS}}}=31.3\substack{+2.1 \\ -2.0}\%$ and a slightly higher value at $z \approx 0.2$ ($f_{\mathrm{B/P,\:\mathrm{det,\:COSMOS}}}=37.8\substack{+5.4 \\ -5.1} \%$). For higher redshifts, $f_{\rm B/P, det}$ declines steeply, reaching $0\substack{+3.6 \\  -0.0} \%$ at $z \approx 1$. A higher $f_{\mathrm{B/P,\:det}}$ for the first COSMOS bin compared to SDSS (considering the uncertainties) is expected, because at redshifts $z<0.3$ \textit{HST} has better spatial resolution compared to the mean spatial resolution of SDSS. Nevertheless, it is a noteworthy aspect that the SDSS and COSMOS B/P bulge fractions at $z<0.3$ are similar. Even though the classifications were done on different imaging, the continuity at the lowest redshift suggests that the observed evolution of the B/P bulge fraction at higher $z$ is real. 

A general trend of decreasing $f_{\mathrm{B/P,\:det}}$ with redshift is what one would expect if the resolution of the images affects the identification of B/P bulges. Since the SDSS images were selected to have roughly the same spread of spatial resolution as the COSMOS data, one would expect $f_{\mathrm{B/P,\:\mathrm{det,\:SDSS}}}$ to be approximately the mean of all $f_{\mathrm{B/P,\:\mathrm{det,\:COSMOS}}}$. However, the mean $f_{\mathrm{B/P,\:\mathrm{det,\:SDSS}}}=31.3\%$ is larger than the mean $f_{\mathrm{B/P,\:\mathrm{det,\:COSMOS}}}=17\%$ (across all redshifts), which is an indication that the observed trend is a genuine evolution of the B/P bulge fraction with redshift rather than being caused by degrading spatial resolution. 

In Figure \ref{BP_fraction_comparison} we plot the detected fractions of barred galaxies hosting B/P bulges for the two SDSS comparison samples, as a function of the equivalent redshift of the corresponding COSMOS galaxies. This shows how the detection of B/P bulges changes with bandshifting and degrading resolution (in the case of the SDSS sample), and with bandshifting, degrading resolution \textit{and} degrading S/N (in the case of the \textsc{redshifted}-control sample). If resolution and S/N had no effect on the B/P detection, we would expect a flat trend (within uncertainties). In Figure \ref{BP_fraction_comparison}, however, we see a decrease of the fraction of B/P bulges in the control samples at $z > 0.6$; this decrease is stronger for the \textsc{redshifted}-control sample, indicating that degrading S/N does make it harder to detect B/P bulges. Despite this, it is important to note that $f_{\mathrm{B/P,\:\mathrm{det,\:COSMOS}}}$ is consistently lower than the control samples at all redshifts $\ga 0.2$, which indicates that the B/P fraction decreases to higher redshifts more than can be accounted for by the observational biases. We will make further use of this plot in Section \ref{corrected_fractions} to correct the detected COSMOS B/P bulge fractions for the observational biases. 

\subsubsection{Logistic regression}
\label{logistic}

In order to quantify the observed trends and evaluate the statical significance of the findings, we make use of logistic regression\footnote{Implemented with the \texttt{statsmodel} package \citep{statsmodels2010} in \textsc{python}}, modelling the probability, $P$, of a galaxy having a binary property (in this case having a B/P bulge or not) as a function of redshift, $z$ (or the equivalent redshift of the COSMOS galaxies), using a sigmoid function bounded between 0 and 1:
\begin{equation}
P(z)=\frac{1}{1+ \rm{e}^{-(\alpha+\beta z)}}
\label{eqn}
\end{equation}

\noindent{where $\alpha$ and $\beta$ are the intercept and slope coefficients of the logistic fit. We apply this to the COSMOS, SDSS and \textsc{redshifted}-control datasets and determine the fit coefficients, shown in Table \ref{logreg}. We also compute the probabilities $P$ of the galaxies having a B/P bulge with redshift, and overplot them in Figure \ref{BP_fraction_comparison} (with blue dashed lines for COSMOS, red solid line for SDSS and green dashed-dotted line for \textsc{redshifted}-control). The logistic regression takes into account all the data points, as opposed to the values in individual bins, so it is not sensitive to the choice of binning. The $p$-value measures the statistical significance of the slope, under the null hypothesis, i.e. how likely it is that the slope is 0 (if the $p$-value is low we can reject the null hypothesis).  }

The slopes of the logistic fits for the three datasets are all negative. For the original SDSS dataset, the $p$-value is marginal (0.031), but for the redshifted-control dataset and the COSMOS data the slopes are clearly different from zero ($p = 2.5 \times 10^{-4}$ and $2.2 \times 10^{-9}$, respectively). More importantly, the \textit{values} of the slopes are clearly different, with the slope being steepest for the COSMOS data ($\beta = -3.89 \pm 0.73$ for COSMOS vs. $-1.78 \pm 0.50$ for the \textsc{redshifted}-control dataset). This means that while S/N effects can \textit{partly} account for the observed decrease in $f_{\mathrm{B/P,\:det}}$ to higher redshift in the COSMOS dataset, they cannot explain all of it: the decrease in $f_{\mathrm{B/P}}$ with increasing redshifts is a real effect.

\begin{figure}
\centering
 \includegraphics[width=1\columnwidth]{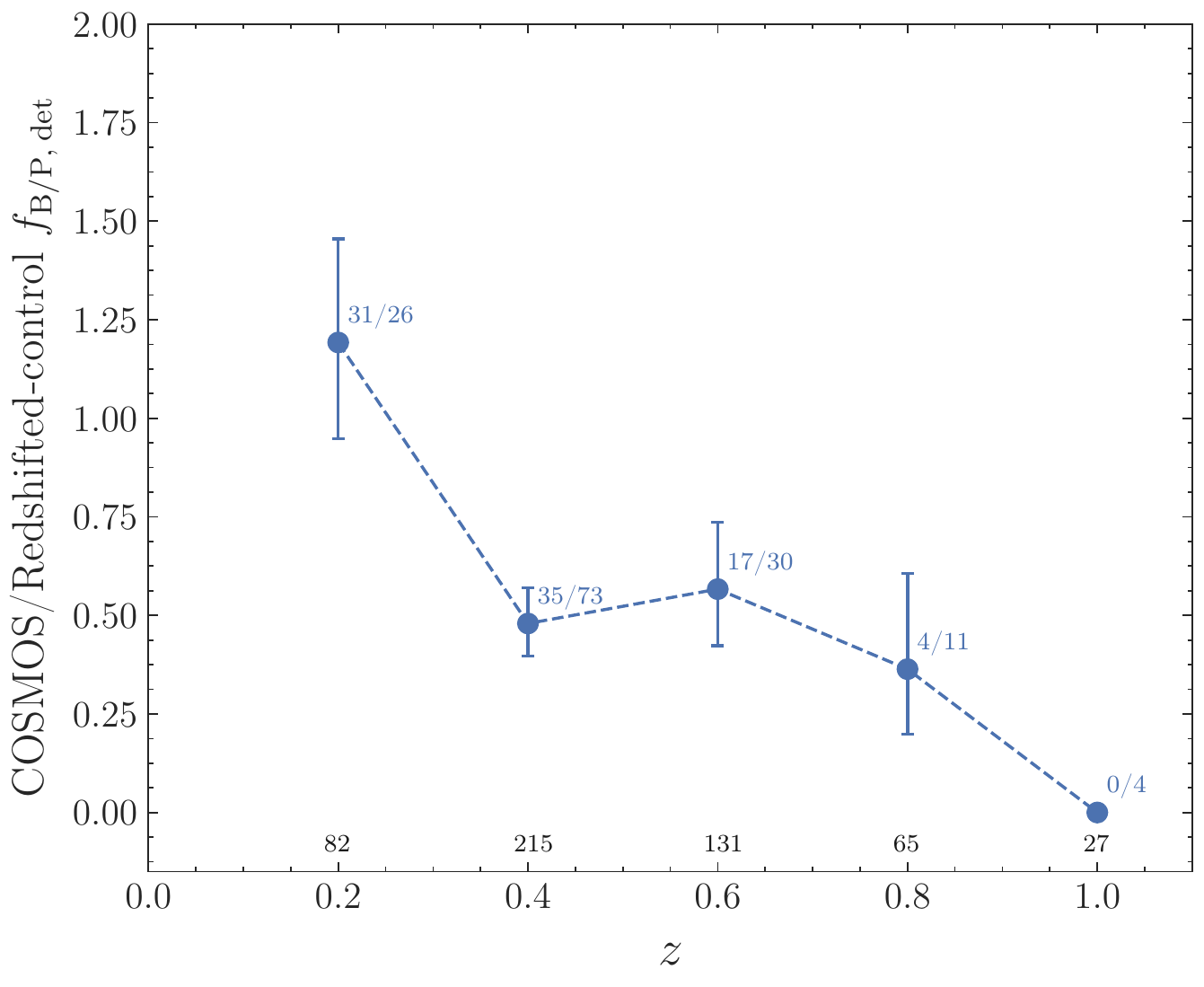}
 \caption{The ratio of detected B/P fractions (COSMOS versus \textsc{redshifted}-control SDSS galaxies) as a function of redshift. The individual points are the ratio of $f_{\mathrm{B/P,\:det}}$ for the COSMOS galaxies to $f_{\mathrm{B/P,\:det}}$ for the \textsc{redshifted}-control SDSS galaxies (i.e. SDSS galaxies with redshift-equivalent spatial resolutions). Since each bin has the same number of COSMOS and SDSS galaxies, this is equivalent to number ratio of detected B/P bulges in the two subsamples. The blue numbers indicate the number of detected B/P bulges in the COSMOS (first number) and SDSS (second number) subsamples; the black numbers are the total number of barred galaxies in each subsample (COSMOS or SDSS).} 
 \label{COSMOS_relative}
\end{figure}

Figure \ref{COSMOS_relative} shows the ratio between the observed B/P fraction in the COSMOS sample and the observed fraction in the \textsc{redshifted}-control sample. At $z = 0.2$ the ratio is $\sim 1$, or perhaps slightly larger (if it is genuinely larger, this is probably due to the fact that at the lowest redshifts, the COSMOS data tends to have better resolution than even the best SDSS images). This ratio drops to $\sim 0.5$ by $z = 0.4$, $\sim 0.4$ at $z = 0.8$, and $\sim 0$ at $z = 1$. This is, again, a clear indication that the fraction of B/P bulges decreases to higher redshift, \textit{after} accounting for bandshifting, changing resolution, and changing S/N in the COSMOS dataset. 

To summarize, we find an evolution of the detected fraction of B/P bulges in barred galaxies from $f_{\mathrm{B/P,\:det}}\approx40\%$ at $z=0$ to $f_{\mathrm{B/P,\:det}}\approx0\%$ at $z=1$. Present day B/P bulge fractions are reached at $z\sim0.25$ (3 Gyr ago), then the B/P bulge fraction declines with increasing redshift to $z\sim1$. The steeper decrease between the $z=0.2$ and $z=0.4$ redshift bins (as shown by the blue data points) appears to be stochastic and due to the choice of threshold for the B/P classification and binning, as it does not appear as evidently in Figure \ref{BP_probability_vs_redshift}. These B/P fractions represent lower limits of the true fraction of B/P bulges in galaxies, since the galaxy samples were selected having all possible bar-disc relative orientations, even with the unfavourable $\Delta\mathrm{PA}_{\mathrm{bar}}\geq60^{\circ}$ to identify B/P bulges using the box and offset spurs signatures in the isophotes. We correct for this effect, by measuring the probability of having galaxies with unfavourable orientations in the sample in Section \ref{corrected_fractions}.  Nevertheless, the B/P fractions reported in this section are the ones that one would measure when inspecting a large random sample of massive local and high-redshift barred galaxies.

\begin{table}
\begin{tabular}{|c|c|c|c|c|l|}
\hline
Dataset                      & $\alpha$               & $\sigma_{\alpha}$         & $\beta$                    & $\sigma_{\beta}$          & $p$-value            \\ \hline
\multicolumn{1}{|c|}{COSMOS} & 0.11                       & 0.31                      & -3.89                      & 0.73                      & 2.2$\times10^{-9}$ \\ \hline
\multicolumn{1}{|c|}{SDSS (control)}   & -0.31                      & 0.24                      & -0.98                      & 0.46                      & 0.031              \\ \hline
$\textsc{redshifted}$-control   & \multicolumn{1}{l|}{-0.12} & \multicolumn{1}{l|}{0.25} & \multicolumn{1}{l|}{-1.78} & \multicolumn{1}{l|}{0.50} & 2.5$\times10^{-4}$ \\ \hline
\end{tabular}
\caption{Logistic fit coefficients, based on Eq. \ref{eqn}: the probability of a barred galaxy having a B/P bulge as a function of redshift for the COSMOS dataset, and equivalent redshift of the COSMOS galaxies for the SDSS datasets. $\alpha$ is the intercept value, at $z=0$, $P(z=0)=\frac{1}{1+\rm{e}^{-(\alpha)}}$ , and $\beta$ is the slope value for the fit, while $\sigma$ denotes the standard error on the parameters. The $p$-value represents the statistical significance that the slope $\beta$ differs from 0.}
\label{logreg}
\end{table}

\subsection{Boxy/peanut bulge fraction with mass}
\label{bp_fraction_mass}

\begin{figure}
\centering
 \includegraphics[width=\columnwidth]{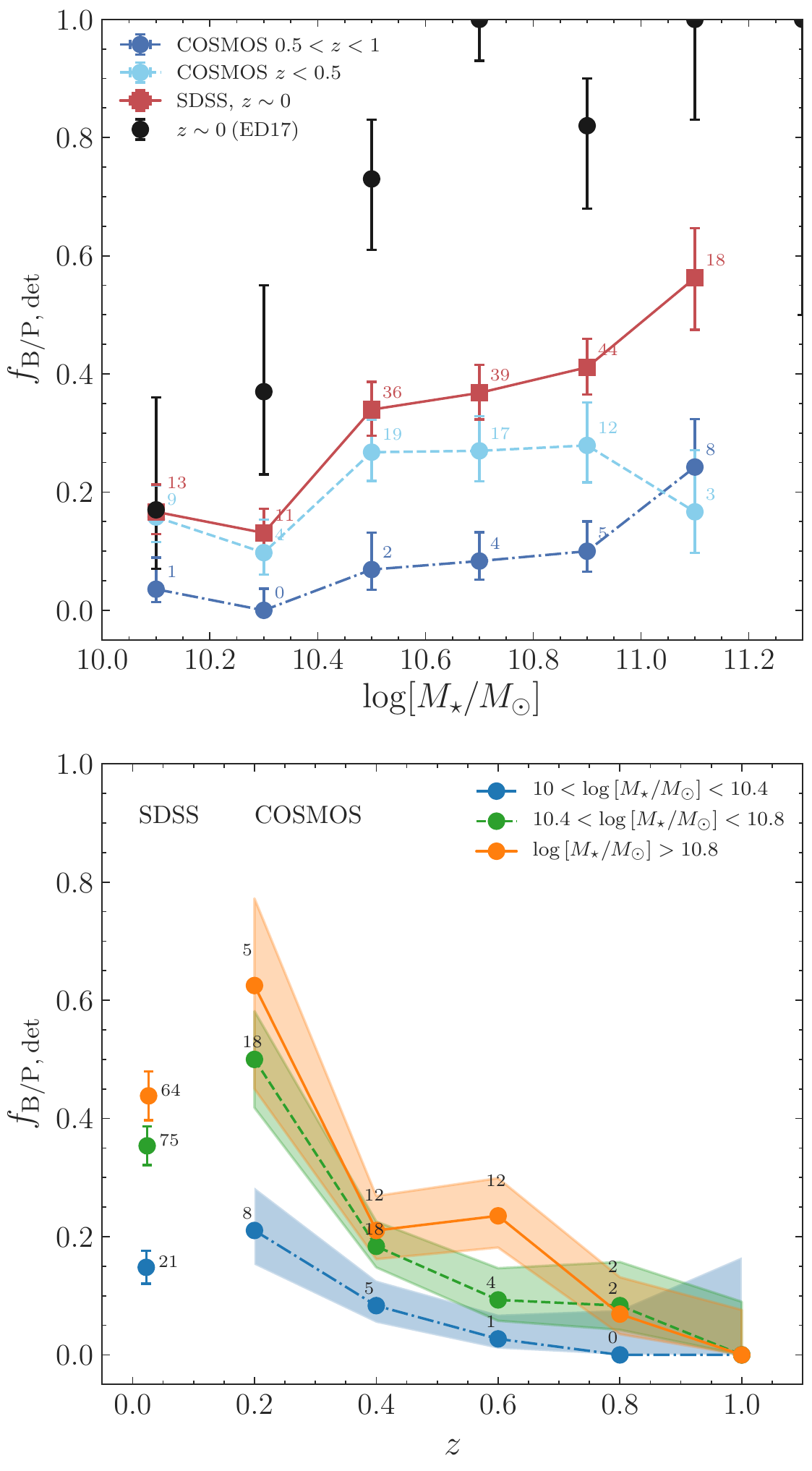}
 \caption[B/P bulge fraction with stellar mass for COSMOS and SDSS galaxies]{\textit{Top panel}: The trend of the B/P bulge fraction with stellar mass for SDSS galaxies (red) and COSMOS galaxies in two redshift bins (light blue and dark blue). The black points show the observed local fraction from \citet{Erwin2017}, for galaxies preferentially selected with ideal orientation to identify B/P bulges. The numbers illustrate the number of B/P bulges galaxies in each bin. \textit{Bottom panel}: The redshift evolution of $f_{\mathrm{B/P,\:det}}$ for three COSMOS mass bins and the value of the local B/P bulge fraction for the SDSS galaxies, for the same three mass bins. The points show the $f_{\mathrm{B/P}}$ in each redshift bin, and the shaded area shows 68\% ($1\sigma$) confidence interval. The numbers represent the total number of B/P bulges in each bin.}
 \label{HST_SDSS_BP_vs_mass}
\end{figure}

Figure \ref{HST_SDSS_BP_vs_mass} (top panel) shows the trend of $f_{\mathrm{B/P,\:det}}$ with stellar mass for the SDSS and COSMOS samples (split into two redshift bins), in comparison with the trend found by ED17 for their ideally selected local sample. The measured $f_{\mathrm{B/P,\:det}}$ increases with stellar mass for the local SDSS data, similar to the trend observed in ED17 and \citet{Li2017}. As also found by ED17, there is a strong change in the B/P bulge fraction at $M_{\star}\sim10^{10.3}-10^{10.5} M_{\odot}$, where $f_{\mathrm{B/P,\:det}}$ doubles. The important new result is the $f_{\mathrm{B/P,\:det}}$ dependence on stellar mass for the higher redshift COSMOS galaxies, shown in Figure \ref{HST_SDSS_BP_vs_mass} for two redshift bins ($z<0.5$ and $0.5<z<1$). The observed dependency of $f_{\mathrm{B/P,\:det}}$ on stellar mass for the COSMOS $z<0.5$ data is similar to the one for the local SDSS data \textendash $ $ $f_{\mathrm{B/P,\:det}}$ increases with stellar mass, also showing a turnover at the same mass threshold. There is some evidence of a transition in $f_{\rm{B/P},\:\mathrm{det,\:COSMOS}}$ at the same stellar mass for the $0.5<z<1$ COSMOS galaxies, although this is based on a smaller number of galaxies. Additionally, as shown earlier, the decrease of the B/P bulge fraction with redshift is also noticeable. In the $0.5<z<1$ redshift bin, $f_{\mathrm{B/P,\:det}}$ is three times lower than for the local barred galaxies. Only the local SDSS and highest COSMOS bin show a `jump' in $f_{\mathrm{B/P}}$ at the largest masses $M_{\star}\sim10^{11} M_{\odot}$, although this mass bin at mid-redshifts in COSMOS contains the fewest galaxies, so it might be affected by small number statistics.

\begin{figure*}
\centering
 \includegraphics[width=1\textwidth]{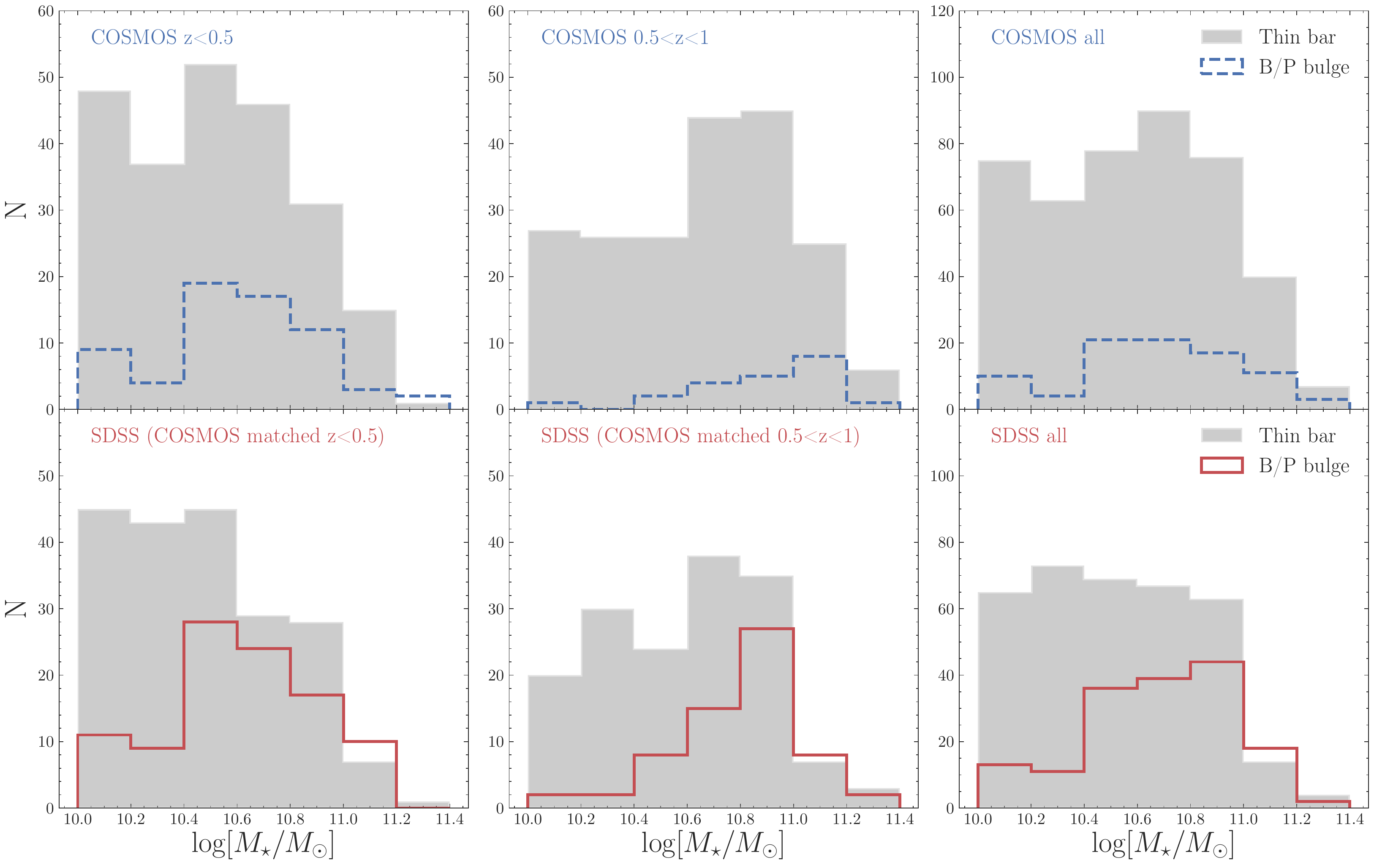}
 \caption{Distribution of stellar masses for galaxies classified as having B/P bulges and thin bars in the COSMOS sample, split into two redshift bins, $z<0.5$ (\textit{top-left panel}) and $0.5<z<1$ (\textit{top-middle panel}), as well as for the entire COSMOS sample  (\textit{top-right panel}). For comparison, the stellar mass distribution of the B/P bulges and thin bars in the matched original SDSS dataset, split into the same redshift bins of the corresponding COSMOS galaxies  (\textit{bottom panels}). The stellar mass distributions of B/P bulges and thin bars are remarkably similar for the COSMOS and SDSS galaxies, within the same redshift bin. The \textit{right panels} show the distribution of B/P bulges and thin bars in the entire COSMOS and SDSS datasets, respectively.}
 \label{HST_SDSS_mass_distrib}
\end{figure*}

Compared to the B/P bulge fraction found in ED17, the local B/P bulge fraction in SDSS data found in this paper is $\sim$ two times lower: $f_{\mathrm{B/P,\:det}}\sim15\%$ at $M_{\star}<10^{10.4} M_{\odot}$ and $f_{\mathrm{B/P,\:det}}\sim40\%$ at $M_{\star}>10^{10.5} M_{\odot}$. The galaxies in ED17 are preferentially selected with favourable inclinations ($i=40^\circ-70^\circ$) and bar position angle within $60^\circ$ of the major axis of the galaxy ($\Delta  \mathrm{PA_{bar}}\leq60^\circ$) to maximize the possibility of B/P bulge identification. In this paper, some COSMOS and SDSS galaxies have low inclinations, $i\lesssim40^\circ$ (see Figure \ref{HST_SDSS_mass_resolution_pbar}), and all possible relative disc-bar orientations, which explains the lower B/P measured fraction. See Section \ref{corrected_fractions} for our attempt to correct for this effect.

The bottom panel of Figure \ref{HST_SDSS_BP_vs_mass} shows the redshift evolution of $f_{\mathrm{B/P,\:det}}$ for three mass bins. These all show decreasing B/P bulge fractions with redshift. The B/P fractions in the first \textit{HST} bin are larger than those for SDSS; this is probably because these COSMOS galaxies have higher resolution than the mean of the SDSS sample. The highest mass bin is noisier because of the smaller number of galaxies. 

To investigate the mass dependence of B/P bulges with redshift, we plot the distribution of stellar masses of the classified B/P bulges and thin bars for the COSMOS galaxies, in two redshift bins, $z<0.5$ (Figure \ref{HST_SDSS_mass_distrib}
top-left panel) and $0.5<z<1$ (Figure \ref{HST_SDSS_mass_distrib} top-middle panel). There is an evident difference between the stellar mass distributions for both thin bars and B/P bulges in the two COSMOS redshift bins (a two-sample Kolmogorov-Smirnov K-S test for B/P bulges gives $k=0.36$, $p_{\mathrm{KS}}=3\times10^{-4}$). However, this is due to the relatively small area and large volume probed by the \textit{HST} COSMOS survey: barred galaxies at $z>0.5$ have on average higher masses by $\sim0.15$ dex than those at $z<0.5$. With our matching approach, we can compare the stellar mass distribution of B/P bulges in COSMOS with that of the resolution and mass-matched SDSS sample in the bottom panels of Figure \ref{HST_SDSS_mass_distrib}, split into the same redshift bins of the corresponding COSMOS galaxies. The mass distribution of B/P bulges in COSMOS is very similar to that of the SDSS control galaxies, when separated in the two redshift bins (K-S test for the mass distribution of B/P bulges in COSMOS $z<0.5$ and SDSS control sample gives $k=0.06$, $p_{\mathrm{KS}}=0.99$, and between COSMOS $0.5<z<1$ and SDSS control sample gives $k=0.28$, $p_{\mathrm{KS}}=0.15$). The similarity in the stellar mass distribution of B/P bulges across all redshifts suggests that stellar mass, or another galaxy parameter with a strong dependence on stellar mass, is the fundamental property that determines the bars to vertically thicken and form B/P bulges. 

\subsection{Corrected B/P bulge fractions}
\label{corrected_fractions}

\begin{figure}
\centering
 \includegraphics[width=1.1\columnwidth]{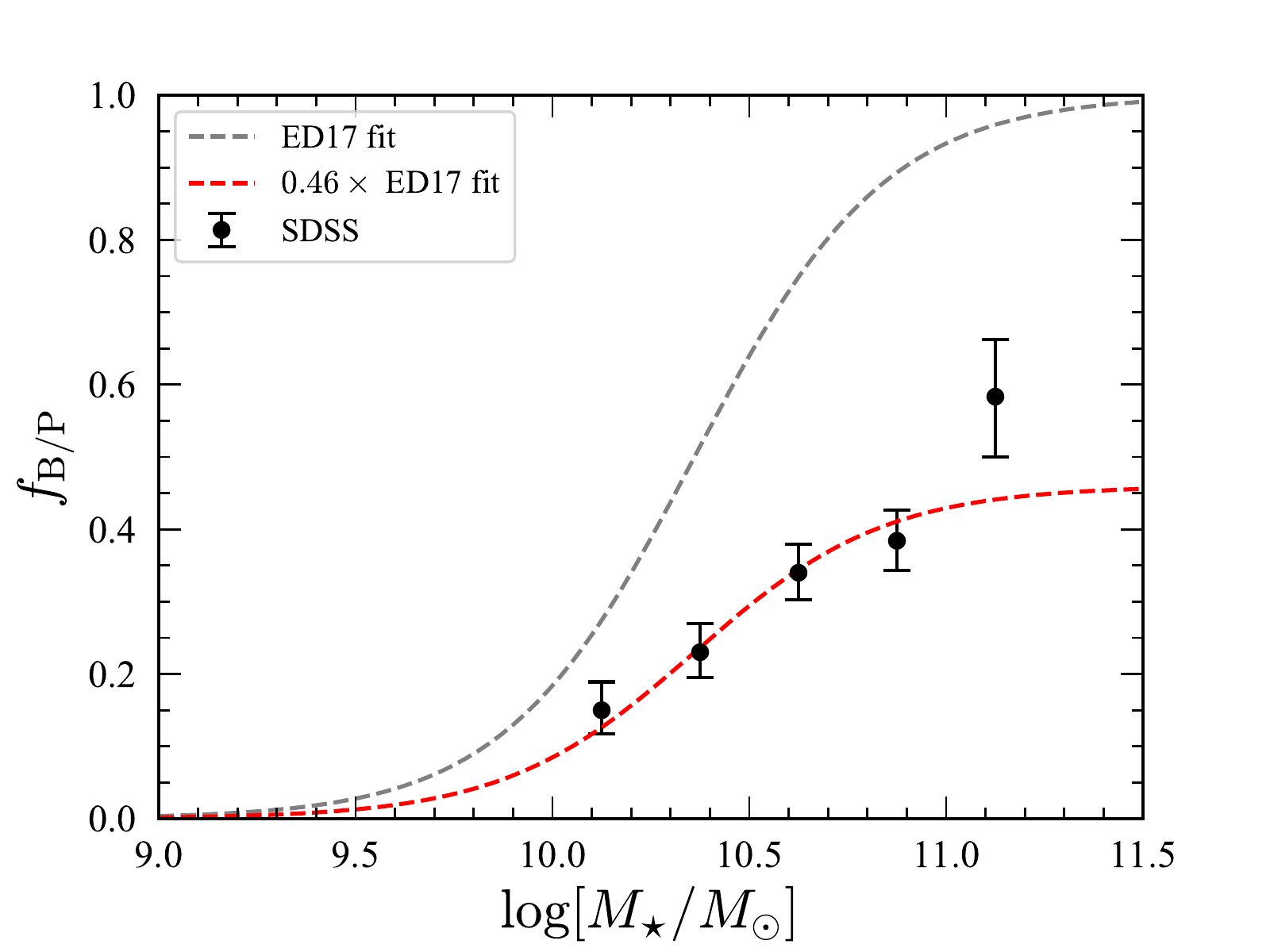}
 \caption{Detected frequency of B/P bulges in SDSS dataset (black points), along with the original logistic fit for the ED17 local sample (gray dashed curve) and the same multiplied by our estimated detection efficiency for the SDSS dataset of 46\% (red dashed curve).} 
 \label{f_BP_SDSS_ED17}
\end{figure}

As noted previously, our approach in identifying B/P bulges uses a method which works best for moderately inclined galaxies ($i \approx 40$--$70\degr$) where the bar is not too close to the minor axis (e.g., bar position angle $\la60\degr$ away from the major axis in the galaxy plane); see the discussion in \citet{Erwin2013}. The SDSS and COSMOS images we use are of galaxies selected to not be ``edge-on'', but which otherwise are not restricted by inclination or bar position angle, and so will include galaxies too close to face-on and/or with bars too close to the minor axis to allow clear identification of B/P bulges. 

This means that we will systematically \textit{underestimate} the true B/P frequency in barred galaxies due to geometric/orientation effects. Since these effects should apply equally to \textit{all} the galaxies, regardless of source, redshift, spatial resolution, etc., they should not affect our ability to detect changes in the \textit{relative} B/P frequencies as a function of redshift. Still, we would like to have some idea of what the `true' B/P frequencies are as a function of redshift, and so we attempt to estimate how often we miss detecting B/P bulges due to orientation/projection effects.

Assuming that galaxies are randomly oriented in space, the probability of a galaxy having inclination $i$ is proportional to $\sin i$. For a sample with a maximum inclination $i_{\rm max}$, the probability of a ``low-inclination'' galaxy with $i \leq i_{\rm min}$ is

\begin{equation}
P_{\rm low inc} = \int_{0}^{i_{\rm min}} \sin i \, \mathrm{d}i \Big/ \int_{0}^{i_{\rm max}} \sin i \, \mathrm{d}i \; = \; \frac{1 - \cos(i_{\rm min})}{1 - \cos(i_{\rm max})} .
\end{equation}

The probability of a barred galaxy having its bar within $\Delta$PA degrees of the galaxy's minor axis ($P_{\rm bad PA}$) is just $\Delta$PA$/90$. If we assume that B/P bulges cannot be seen for $i < i_{\rm min}$ and that they cannot be seen in higher inclination galaxies when the bar position-angle is too close to the minor axis, then the probability of galaxy having an unfavourable orientation is 

\begin{equation}
P_{\rm low inc} \; + \; (1 - P_{\rm low inc}) P_{\rm bad PA} .
\end{equation}

For values of $i_{\rm min} = 30$\degr-- 40\degr, $i_{\rm max} = 65$\degr{}--70\degr{} (e.g., Figure \ref{HST_SDSS_mass_resolution_pbar}), and $\Delta$PA $= 20\degr- 30\degr,$ we find that $\sim40$--60\% of the galaxies will have bad orientations, making it difficult or impossible to identify a B/P bulge. So we can expect to miss $\sim$ half of the B/P bulges in our samples.

We can also estimate how many B/P bulges we are missing by comparing the results of our analysis of the SDSS sample with the previous analysis of local galaxies by ED17. The latter study was explicitly restricted to barred galaxies with $i = 40$\degr--70\degr{} and bar PA $< 60\degr$ away from the major axis (in the plane of the disc), in order to maximize the probability of correctly identifying B/P bulges. In addition, the analysis was done using primarily near-infrared images, which minimizes the possibility of confusion due to dust and strong star formation, so it should represent a close approximation to the true frequency of B/P bulges. As part of that analysis, ED17 found that B/P frequency was a very strong function of stellar mass (see their Section~4 and Figure~5). Assuming the latter relation holds (and Figure \ref{HST_SDSS_BP_vs_mass} shows evidence that it does), we can use the ED17 logistic regression result ($P_{\rm B/P}(\log M_{\star})$; parameters from their table~2) to estimate how many of the barred SDSS galaxies \textit{should} have B/P bulges, and compare that to the number we actually identified. The prediction is that 66.2\% of the SDSS sample should possess B/P bulges, versus an observed frequency of 31\%. This suggests that our detection efficiency is $\sim 46$\%, which is entirely consistent with the geometric arguments above.

Figure \ref{f_BP_SDSS_ED17} shows the observed frequency of B/P bulges in the SDSS dataset as a function of stellar mass, along with the ED17 logistic curve in black and 0.46 times that same curve in red. The latter is a remarkably good match to the observed frequencies, except in the very highest mass bin, where the numbers are low and the uncertainty is high.

The analysis above suggests that we are able to detect $\sim$ half of the B/P bulges in our sample. Therefore, the `true' B/P fractions should be $\sim$ two times larger, for both SDSS and COSMOS, since both samples have similar inclinations and were selected to have random bar position angles. We use this factor to estimate a \textit{corrected} fraction of B/P bulges ($f_{\mathrm{B/P,cor}}$), accounting for the presence in our samples of galaxies with low inclinations and bars closer to the galaxies' minor axes. Additionally, we account for the changing resolution, S/N and bandshifting, using the logistic fit to the B/P bulge fraction in the \textsc{redshifted}-control sample, shown in Figure \ref{BP_fraction_comparison} (with a green dash-dotted line). In the absence of any observational bias, the detected B/P fraction in the \textsc{redshifted}-control sample should be constant with the equivalent redshift of the COSMOS galaxy (which is a proxy for the spatial resolution of the SDSS galaxies), and equal to the value at $z=0.2$ (since these galaxies have the best resolution and S/N in the sample), $P_{\mathrm{B/P,\:det, \textsc{redshifted}-control,z=0.2}}=0.38$. We use the $P_{\mathrm{B/P,\:det, \textsc{redshifted}-control,z=0.2}}/P_{\mathrm{B/P,\: det,\:\textsc{redshifted}-control,z}}$ ratio to correct for the degrading resolution and S/N on our classifications, together with the factor of 2 to account for the unfavourable orientations, in order to calculate the \textit{corrected} B/P bulge fractions

\begin{equation}
f_{\mathrm{B/P,\:cor}}(z)=2\times f_{\mathrm{B/P,\:det}}(z) \times \dfrac{{P_{\mathrm{B/P,\:det, \textsc{redshifted}-control, z=0.2}}}}{P_{\mathrm{B/P,\: det,\:\textsc{redshifted}-control,z}}}.
\label{correction}
\end{equation}

\begin{figure}
\centering
 \includegraphics[width=1\columnwidth]{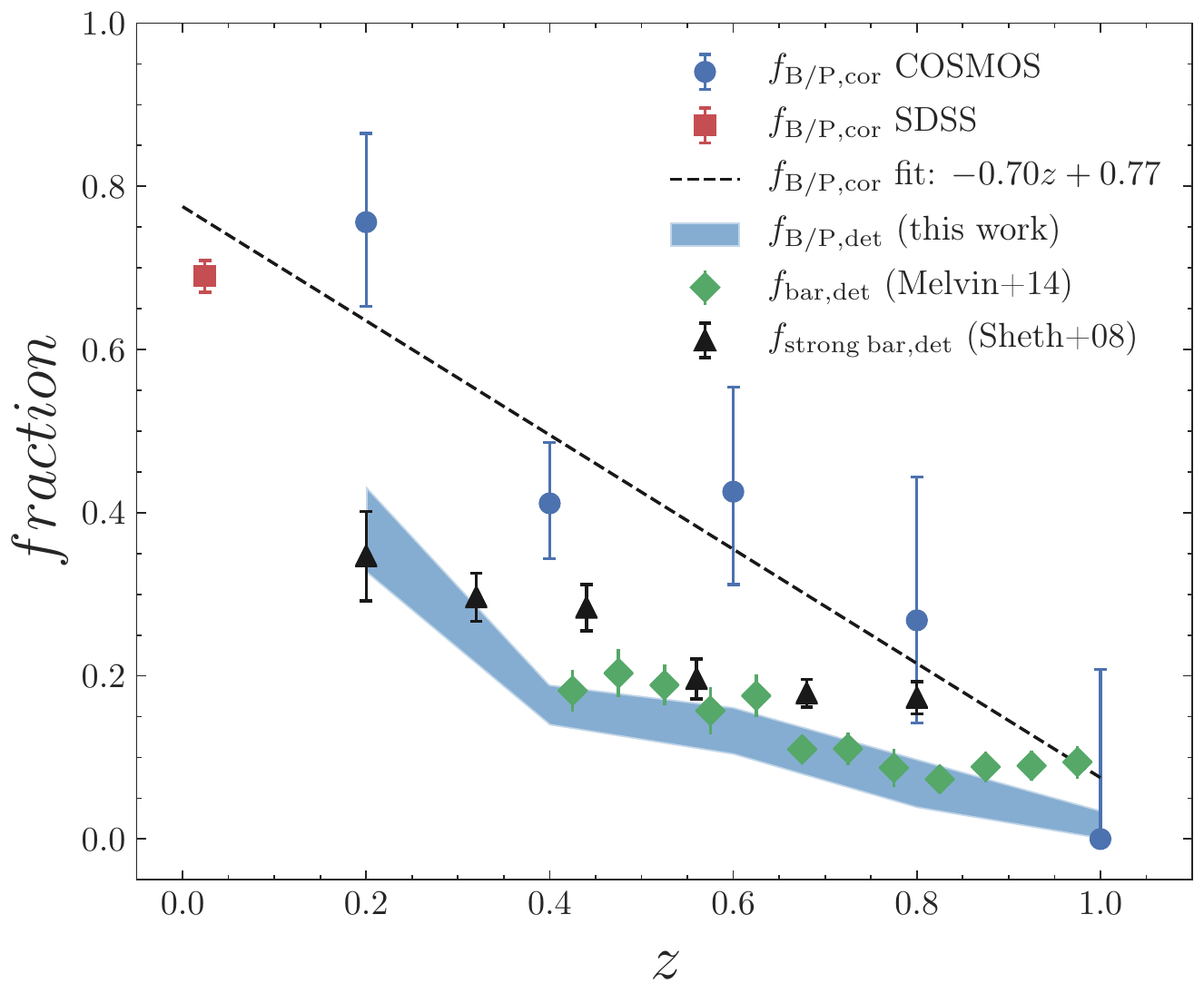}
 \caption{Fraction of barred galaxies with B/P bulges corrected for unfavourable inclinations and bar orientations in our sample, as well as changing resolution, S/N and bandshifting, using Eq. \ref{correction} (blue circles for COSMOS and red squares for SDSS). The trend declines with increasing redshift, showing the genuine evolution in the B/P fraction. The dashed line shows a linear fit to the mean, binned, corrected B/P fractions, from $z\approx0$ to $z=1$. For comparison, the detected B/P bulge fraction ($f_{\mathrm{B/P,\:det}}$) is shown with the blue band. The detected fractions of disc galaxies with strong bars in COSMOS from \citet{Sheth2008} (green diamonds) and \citet{Melvin2014} (black triangles) are also represented. The error bars in $f_{\mathrm{B/P,\:cor}}$ show the combined uncertainties on $f_{\mathrm{B/P,\:det}}(z)$ ($1\sigma$ limits in the binomial confidence interval, shown in Figure \ref{BP_vs_redshift}) with the uncertainties in the coefficients of the logistic regression fitted to the \textsc{redshifted}-control sample (parameters shown in Table \ref{logreg}) and propagated using Eq. \ref{correction}. } 
 \label{f_BP_corrected}
\end{figure}

The final, corrected B/P bulge fractions are shown in Figure \ref{f_BP_corrected}. Applying the correction in Eq. \ref{correction}, the fraction of B/P bulges in COSMOS declines from $\sim75\%$ (at $z=0.2$), to $\sim40\%$ at $z=0.4$ and $z=0.6$ and $\sim30\%$ at $z=0.8$. 

In addition, we corrected the $z\approx0$ B/P bulge fraction in SDSS for the detection efficiency. For this, we considered only the highest resolution and S/N SDSS images in our sample. The detected B/P fraction for these galaxies corresponds to the `control SDSS' B/P fraction at (equivalent \textit{HST} redshift) $z=0.2$ in Figure \ref{BP_fraction_comparison}, $f_{\mathrm{B/P,\:det}}=0.345$, which is multiplied by the geometric factor of 2, yielding a corrected SDSS B/P bulge fraction of $\sim69\%$\footnote{The corrected B/P bulge fraction in the local Universe agrees well with other studies. For example, investigating B/P bulges in the $i$-band images of a sample of $\sim1,300$ nearby edge-on galaxies from SDSS, \citet{Yoshino2015} found a B/P bulge fraction of $22\%$. This is the fraction of B/P bulges in \textit{all} edge-on galaxies they studied, which contains both barred and non-barred galaxies. Since they also estimated the fraction of barred galaxies in face-on galaxies in their sample, $f_{\mathrm{bar}}\approx33\%$, this implies a total fraction of $\sim66\%$ of SDSS barred galaxies hosting B/P bulges, in good agreement with our study.}. We use a linear regression to fit the corrected fractions with redshift. The linear fit implies a small but non-zero ($\lesssim10\%$) `real' B/P bulge fraction at $z=1$. 

In Figure \ref{f_BP_corrected} we also plotted, for comparison, the \textit{detected} B/P bulge fraction from Figure \ref{BP_vs_redshift} and the fraction of barred disc galaxies in \textit{HST} COSMOS, based on two independent studies. \citet{Sheth2008} identified bars based on ellipse fits and expert visual inspection, finding a decreasing fraction of strong bars from $f_\mathrm{bar}=35\%$ at $z=0.2$ to $f_\mathrm{bar}\approx15\%$ at $z=0.8$. Using a similar method for selecting strongly barred galaxies as in this work (based on Galaxy Zoo classifications), \citet{Melvin2014} also found a decreasing bar fraction from $f_\mathrm{bar}=22\%$ at $z=0.4$ to $f_\mathrm{bar}\approx11\%$ at $z=1$. The fraction of barred galaxies is relatively constant between $z=0.6-1$, $f_\mathrm{bar}\approx10\%$, after which it increases linearly to $z=0.2$ with a slope that appears similar to the slope of the detected fraction of B/P bulges. Nevertheless, since the detection of bars and the detection of B/P bulges with the method described in this paper suffer from different observational biases, we are reluctant to interpret the observed similarity of the two trends.

\section{Discussion}
\label{discussion}

We have investigated the redshift evolution of boxy/peanut bulges with data from the $\textit{HST}$ COSMOS survey and local SDSS galaxies. For barred galaxies with $M_{\star} > 10^{10}\:M_{\odot}$, we find a detected B/P fraction of $31.3^{+2.1}_{-2.0}$\% at $z \approx 0$, decreasing to $6.2^{+3.6}_{-2.4}$\% at $z = 0.8$ and to $0^{+3.6}_{-0.0}$\%  at $z = 1$. Corrected for observational biases, low galaxy inclinations and unfavourable bar orientations to detect B/P bulges in our sample, the fraction of B/P bulges evolves from $\sim$70\% at $z=0$ to $\lesssim10\%$ at $z=1$. This suggests that there is a time delay between when the bars form and when they vertically thicken, otherwise we would observe a constant fraction of B/P bulges. We find that stellar mass, in addition to redshift, is an important determinant of whether a bar is boxy/peanut shaped or not. There is a strong increase in the B/P fraction at a stellar mass of $M_{\star}\gtrsim10^{10.4}\:M_{\odot}$, where the fraction of barred galaxies hosting B/P bulges $\sim$ doubles, in both low-redshift and high-redshift datasets. 

\begin{figure*}
\centering
 \includegraphics[width=1\textwidth]{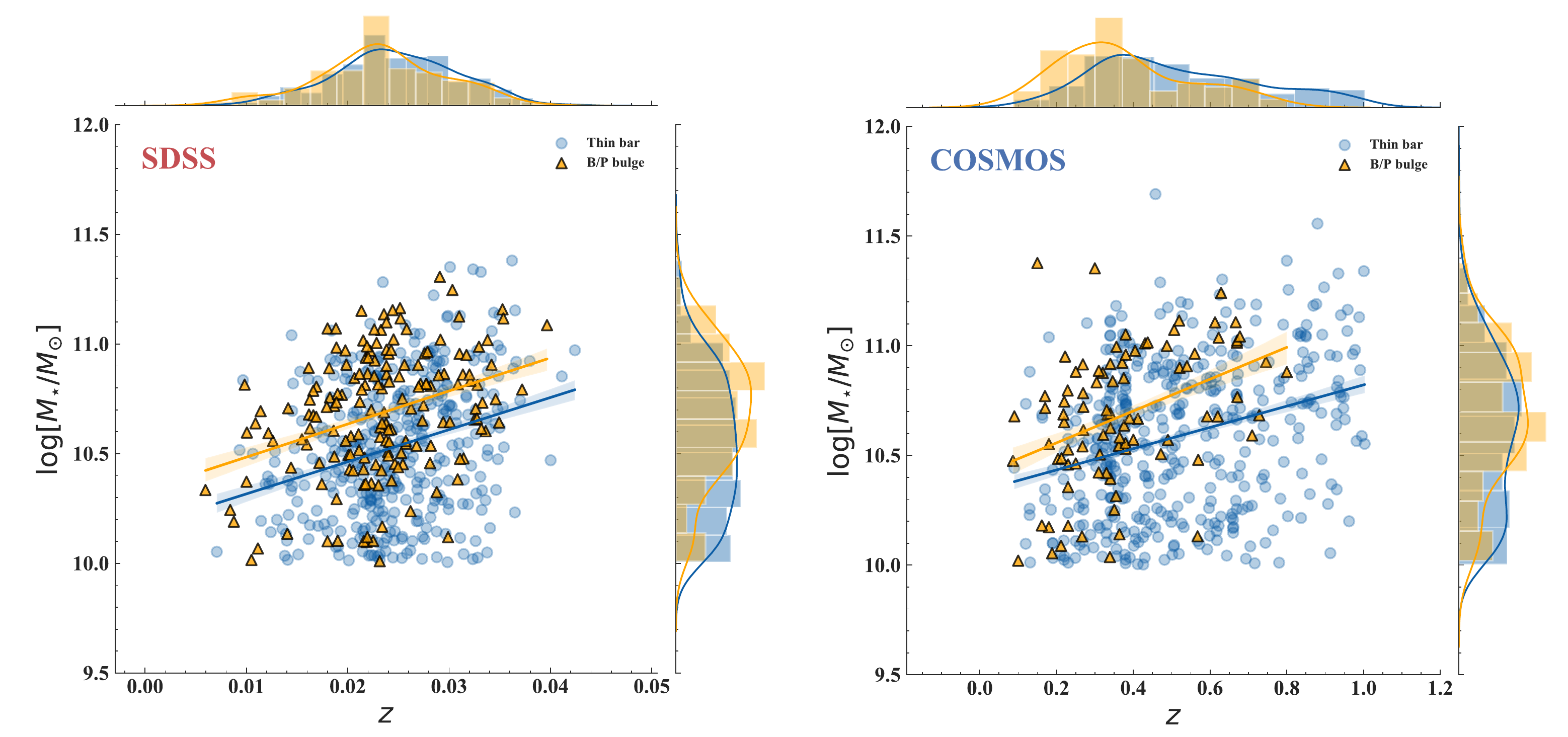}
 \caption{Mass-redshift plots for the local SDSS sample (\textit{left}) and higher redshift COSMOS sample (\textit{right}), split into B/P bulges (triangles) and thin bars (circles), with linear regressions to the two datasets (with the shaded area showing the 1$\sigma$ confidence interval). The marginal histograms show the mass and redshift distributions for the datasets.}
 \label{mass_redshift_HST_SDSS}
\end{figure*}

To summarize our findings, we plot the mass-redshift distribution of the B/P bulges identified in this project in both local and high redshift datasets, in comparison to thin bars, in Figure \ref{mass_redshift_HST_SDSS}. Galaxies with detected B/P bulges have higher masses than those without by 0.2 dex, on average, in the case of the SDSS galaxies, and 0.1-0.3 dex, in the case of COSMOS galaxies. It is important to note that, at the same physical resolution ($>0.5$ kpc,  equivalent to $z>0.03$ in SDSS), there is a considerable population of B/P bulges present in the SDSS dataset which is missing from the COSMOS dataset (at $z>0.5$). As shown in Section \ref{results}, the relative absence of B/P bulges at high redshift cannot be accounted for by degrading resolution and S/N
hindering their identification; instead, there is most likely a physical explanation: most barred galaxies at higher redshifts have not yet formed B/P bulges out of their bars. In what follows we discuss the implications of our findings in the context of evolution of barred galaxies. 

One caveat to note is that this study does not take into account the growth in stellar mass of the galaxies since $z=1$. The higher redshift COSMOS galaxies are not the progenitors of the local matched SDSS galaxies considered in this study. Based on the observed stellar mass functions, galaxies have roughly doubled in mass since $z=0.8$ \citep{Wang2010} through a combination of star formation and minor mergers. A galaxy with $M_{\star}=10^{11}\:M_{\odot}$ at $z\approx0$ would have a stellar mass of $M_{\star}\sim10^{10.7}\:M_{\odot}$ at $z=1$ and one with $M_{\star}=10^{10.5}\:M_{\odot}$, $\sim10^{10}\:M_{\odot}$ at $z=1$ \citep{Hill2017}. In the \textsc{Eagle} simulations, galaxies have increased in stellar mass by as much as 0.5 dex, depending if they are active or passive, since $z=1$ \citep{Clauwens2016}. However, the aim of this study was to compare similar mass galaxies, rather than track the B/P formation in possible progenitor galaxies at different redshift which is subject to considerable scatter.

\subsection{When do B/P bulges form?}

In this study, we detect the highest redshift B/P bulges at $z$ = 0.7--0.8 (see Figure \ref{mass_redshift_HST_SDSS}, right panel), suggesting that most B/P bulges that we observe in the local Universe have formed in the last $\sim$ 7 Gyr. Furthermore, the very low corrected fraction of B/P bulges at $z\approx1$, implied from the linear fit in Figure \ref{f_BP_corrected} (and also Appendices \ref{maybe_yes} and  \ref{candels}), suggests that B/P bulges began forming at $z\sim1$, 8 Gyr ago. The local B/P fraction is reached at $z=0.25$ (see e.g. Figure \ref{BP_vs_redshift}), 3 Gyr ago, implying that the vast majority of B/P bulges formed between 8 and 3 Gyr ago. This agrees with integral field spectroscopic observations of NGC 6032 suggesting that the B/P bulge formed via the buckling instability $\sim$ 8 Gyr ago \citep{Perez2017}. Furthemore, spectroscopic observations of barlenses in the CALIFA survey show that the stellar populations of barlenses are similar to the stellar populations in the thin bar region (suggesting a connection between face-on barlenses, inclined B/P bulges and bars), and are, on average, between 4-8 Gyr old \citep{Laurikainen2018}. Since these stars were likely already formed at the time the B/P bulge was formed, this matches our observations of B/P bulges being formed 3-8 Gyr ago. 

Furthermore, observations of high-redshift galaxies in \textit{HST} CANDELS survey place the highest redshift barred galaxies at $z$ = 1.5--2 \citep{Simmons2014}, implying some bars formed $\sim$ 9-10 Gyr ago. Our observations of the highest redshift B/P bulges in \textit{HST} COSMOS at $z$ $\approx$ 0.7--0.8 ($\sim7$ Gyr ago), suggest a time delay between the formation of bars and of B/P bulges of $\sim2$ Gyr, in good agreement with predictions from $N$-body simulations (e.g. \citealt{Martinez-Valpuesta2004,Martinez-Valpuesta2006,Saha2013}).

\subsection{What determines the formation of B/P bulges?}

This study, ED17 and  \citealt{Li2017} all show that there is a strong stellar mass dependence of the fraction of barred galaxies which have B/P bulges. ED17 found that $f_{\mathrm{B/P}}$ increased rapidly over a
relatively narrow transition region of $M_{\star} \sim 10^{10.3}\:M_{\odot}$ to $10^{10.5}\:M_{\odot}$. In this paper's data, we see a similar, albeit weaker, jump for the SDSS galaxies and the lower redshift COSMOS galaxies (Figure \ref{HST_SDSS_BP_vs_mass}). There is even a hint of a jump in $f_{\mathrm{B/P},\:\mathrm{det,\:COSMOS}}$ for the higher redshift galaxies, although this is based on very small numbers and should be considered uncertain.

ED17 investigated whether the fraction of H\textsc{i} gas in galaxies correlates with the presence of a B/P bulge in local galaxies, since some simulations (e.g. \citealt{Debattista2006,Berentzen2007}) suggest that the presence of significant gas can delay the vertical thickening of the bar. They did not find a statistically significant correlation between the amount of gas present in a galaxy and the presence of a B/P bulge, once the strong relation between gas mass and stellar mass is accounted for. However, galaxies currently hosting B/P bulges might have had different amounts of gas in the past, when the B/P bulges formed. Galaxies at higher redshift are expected to have higher gas fractions \citep{Tacconi2010,Tacconi2013}. Investigating the presence of gas in galaxies with and without B/P bulges at higher redshifts might reveal whether gas is truly responsible for suppressing bar buckling; however this was not possible in the current study because there are very few gas measurements in high-redshift barred galaxies. 

The apparent transition mass range of $M_{\star} \sim 10^{10.3}\:M_{\odot}$ to $10^{10.5}\:M_{\odot}$ that we observe for $f_{\mathrm{B/P}}$ roughly corresponds to the mass found by \citet{Kauffmann2003b} at which galaxies in the local Universe change significantly. Lower mass galaxies have young stellar populations, low surface mass densities and are disc dominated. Higher mass galaxies have older stellar populations, high surface mass densities and higher concentrations typical of bulges. \citet{Kauffmann2003b} attributed this sudden change in galaxy properties to feedback process (AGN and supernova feedback) that regulate the growth of galaxies. This mass also corresponds to the characteristic mass of $M_{\star}\approx3\times10^{10}\:M_{\odot}$ in the mass-size ($M_{\star}-r_{e}$) and mass-velocity dispersion ($M_{\star}-\sigma$) relations (shown in Figure 20 of \citealt{Cappellari2016}) $-$ a break in the galaxy scaling relations setting a minimum size for galaxies where the velocity dispersion reaches a maximum. Thus, it is possible that a change in velocity dispersion determines the formation of B/P bulges. The ratio of the vertical to radial velocity dispersion ($\sigma_{z}/\sigma_{r}$) has been shown to be the driver of bar buckling in $N$-body simulations - when the ratio drops below a certain critical threshold because of the increase of $\sigma_{r}$ following bar formation, the disc becomes vertically unstable and the disc undergoes violent buckling instability \citep{Toomre1966,Raha1991,Merritt1994, Debattista2006,Martinez-Valpuesta2006}. 

To test observationally whether the increased vertical velocity dispersion is responsible for suppressing the violent buckling instability, one would need to measure the vertical and radial random motions in galaxies, which is currently difficult to achieve for large samples. However, $\sigma_{z}$ and $\sigma_{r}$ correlate with the vertical and radial scale lengths respectively ($h_{z}$ and $h_{r}$, \citealt{Kruit1988}). Therefore, measuring the scale length ratio of discs ($h_{z}/h_{r}$), or alternatively, measuring the contribution of a thin vs. a thick disc as a function of stellar mass for edge-on galaxies could help determine whether galaxy thickness can indeed cause the buckling of the bar. There is some evidence that the discs in low-mass galaxies are dominated by thick disc stars, in contrast to higher mass galaxies, which have thinner discs \citep{Yoachim2008}. More recently, by fitting the brightness profiles of the thin and thick disc components of edge-on galaxies, \citet{Comeron2018} found that although thick discs are very common in galaxies (in 82\% of their galaxies), the thick discs in low-mass galaxies (with circular velocities $v_{c}<120\:\mathrm{km}\:\mathrm{s}^{-1}$) are comparable in mass to their thin counterparts while the thick discs in high-mass galaxies are typically less massive than the corresponding thin discs. The reported circular velocity corresponds to stellar masses $M_{\star}\approx10^{10}-10^{10.5}\:M_{\odot}$ (based on estimates from the Tully-Fisher relation at $z\approx0$, see e.g. \citealt{Bell2001}); this mass range includes the characteristic mass that we find for the transition in the B/P fraction. We suggest that the transition from a thick to a thin dominating component of the galaxy disc also determines the violent bending instabilities of the bar that subsequently leads to the formation of a B/P bulge. 

Integral field spectroscopic studies of higher redshift galaxies, such as KMOS$^{\mathrm{3D}}$, show that the velocity dispersion of gaseous discs decreases from $z=1$ to $z\approx0$ for galaxies with stellar masses $M_{\star}\approx10^{10.5}\:M_{\odot}$ (see e.g. Figure 10 in \citealt{Wisnioski2015} or Figure 7 in \citealt{Kassin2012}). This implies that higher redshift disc galaxies are thicker compared to local galaxies. Since we find an increase in the fraction of B/P bulges with redshift, from $z\approx1$ to $z\approx0$, this is consistent with the picture that the transition from a thick to a thin dominating discs leads to the formation of B/P bulges. Simulations on the formation of B/P bulges in thin and thick discs, for example \citet{Fragkoudi2017}, have concentrated on the different morphologies and strengths of the vertically thickened bars (stronger, `X'-shaped in thin discs and weaker `boxy'-shaped in thick discs), but not on whether bar buckling can be suppressed in thick discs. Further simulations, exploring different stellar masses, and investigating the correlations between disc thickness and the formation and morphology of B/P bulges are needed to establish if and when bars buckle. Our observations place important constraints on the B/P bulge formation models.

\section{Conclusions}

In this paper, we investigated, for the first time, the presence of boxy/peanut bulges in both the local and higher redshift Universe (to $z\sim1$) using a sample of 520 barred galaxies from SDSS and the \textit{HST} COSMOS survey, with bar classifications from the Galaxy Zoo project. To help correct for observational biases, we matched SDSS and COSMOS galaxies individually by stellar mass, spatial resolution, bar likelihood and rest frame band. To account for the effects of decreasing S/N with increasing redshift in COSMOS galaxies, we also generated noise-added SDSS images to match the S/N of the corresponding COSMOS galaxies. The selected barred galaxies are moderately inclined and face-on, such that the identification of bars and boxy/peanut bulges is possible based on the shapes of the inner isophotes using a method devised by \citet{Erwin2013}, and \citet{Erwin2017}. The galaxy isophotal contours were classified by three authors and the classifications were aggregated to identify barred galaxies hosting B/P bulges, galaxies with thin bars and also galaxies with buckling bars (the latter to be presented in a future paper). We corrected for unfavourable galaxy inclinations and bar orientations in our sample which affects the detection efficiency of B/P bulges. Finally, we determined the local and higher redshift fraction of barred galaxies having B/P bulges. We summarize the findings in this study as follows:

\begin{enumerate}

\item The detected fraction of barred galaxies hosting boxy/peanut bulges decreases steeply with increasing redshift, from $f_{\rm B/P, det} = 31.3^{+2.1}_{-2.0}$\% at $z \approx 0$ for the SDSS sample (and $f_{\rm B/P,\:det} = 37.8^{+5.4}_{-5.1}\%$ for COSMOS galaxies at $z \approx 0.2$) down to $f_{\rm B/P,\:det} = 6.1^{+3.7}_{-2.4}\%$ at $z \approx 0.8$ and $f_{\rm B/P, det} = 0^{+3.6}_{-0.0}\%$ at $z \approx 1$.

\item We detect the highest redshift galaxies with B/P bulges in COSMOS at $z \approx 0.7 - 0.8$, consistent with their formation $\sim$ 7 Gyr ago. 

\item We estimate the fraction of B/P bulges missed due to unfavourable inclinations and bar orientations in our samples to be $\sim 50$\%, independent of redshift. The real B/P bulge fractions are thus probably a factor of $\sim$ 2 larger than the detected fractions listed above (see Section \ref{corrected_fractions}).

\item The identification of B/P bulges, especially at high redshifts, is affected by observational biases such as degrading resolution and signal-to-noise of the galaxy images, as well as morphological bandshifting. 

Accounting for these effects using the artificially redshifted comparison sample selected from SDSS, along with the inclination/bar-orientation effect discussed previously, we find that the estimated true fraction of bars with B/P bulges decreases from $f_{\rm B/P} \approx 70$\% at $z = 0$ (for SDSS) (and $f_{\rm B/P} \approx 75$\% at $z = 0.2$ for COSMOS) to $\sim 40$\% for $z = 0.4$--0.6, $\sim 30$\% for $z = 0.8$, and $\la 10\%$ for $z = 1$. The B/P bulge fractions are consistent with the vast majority of B/P bulges having formed in the last 7--8 Gyr, with a time delay between when the bars form and when they vertically thicken. Modern B/P fractions are reached at $z \approx 0.25$, suggesting that relatively few B/P bulges have formed in the last $\sim 3$ Gyr.

\item The boxy/peanut fraction depends strongly on stellar mass, with $f_{\rm B/P}$ increasing to higher masses. This was already known for local ($z \sim 0$) galaxies, but we find it to be true for high-redshift galaxies as well. The stellar-mass distribution of B/P bulge hosts is similar at low and high redshifts, suggesting the underlying relation does not change greatly with redshift.

\end{enumerate}

These observations give us, for the first time, direct insights into when B/P bulges form. The redshift evolution of the detected fractions and the strong correlation with stellar mass can be used to test the predictions from $N$-body simulations for the different mechanisms that lead to the formation of B/P bulges. Furthermore, studying these external galaxies can help us better understand the formation of the B/P bulge in our own galaxy, the Milky Way \citep{Gonzalez2016}.

The sample of $\sim500$ barred galaxies from the \textit{HST} COSMOS survey used in this study is currently the largest sample of high-redshift galaxies where the spatial resolution and image depth allows us to investigate the presence of B/P bulges up to $z\approx1$. With future large astronomical surveys, such as the \textit{Euclid} Wide survey  \citep{Laureijs2011}, which will image a few orders of magnitude larger area than the \textit{HST} Legacy Surveys ($15,000\:\mathrm{deg}^2$), but with a slightly lower spatial resolution ($\approx0.2^{\prime\prime}$ with the VIS instrument), it will be possible to study millions of barred galaxies, thus significantly improving the statistics in this study. Furthermore, with the high-resolution infrared imaging and spectroscopy of \textit{JWST} it will be possible to study in detail a selected subsample of galaxies to pinpoint the process in which B/P bulges form.

\appendix

\section{Classification bias}
\label{maybe_yes}

In this paper, three authors classified galaxies as one of `No bars', `No', `Maybe', `Probably' and `Definitely' B/P bulges. These were then aggregated into positive or negative B/P classification based on the total score assigned to each classification (the individual scores were -1, 0, 1, 2, and 3, from `No bar' to `Definitely). A threshold of 3 was selected as the threshold between a galaxy being classified as having a \textit{thin bar} or a \textit{B/P bulge}. This choice of threshold is nevertheless somewhat arbitrary, so one needs to check for possible biases it introduces in the results. In what follows, we compute a B/P bulge `likelihood' ($p_{\mathrm{B/P,\:det}}$) for each galaxy, by computing the probability for each B/P bulge classification (`No', `Maybe', `Probably'  and `Definitely') and adding them together with weights of 0\%, 25\%, 75\% and 100\%, respectively. The `No bar' and `No' B/P bulge classifications are given weights of 0\%, and do not count in the $p_{\mathrm{B/P,\:det}}$ calculation. The B/P `likelihood' is thus:

\begin{equation}
p_{\mathrm{B/P,\:det}}=0.25\times p_{\mathrm{Maybe}}+0.75\times p_{\mathrm{Probably}}+p_{\mathrm{Definitely}}.
\label{BP_likelihood}
\end{equation}

\begin{figure}
\centering
 \includegraphics[width=\columnwidth]{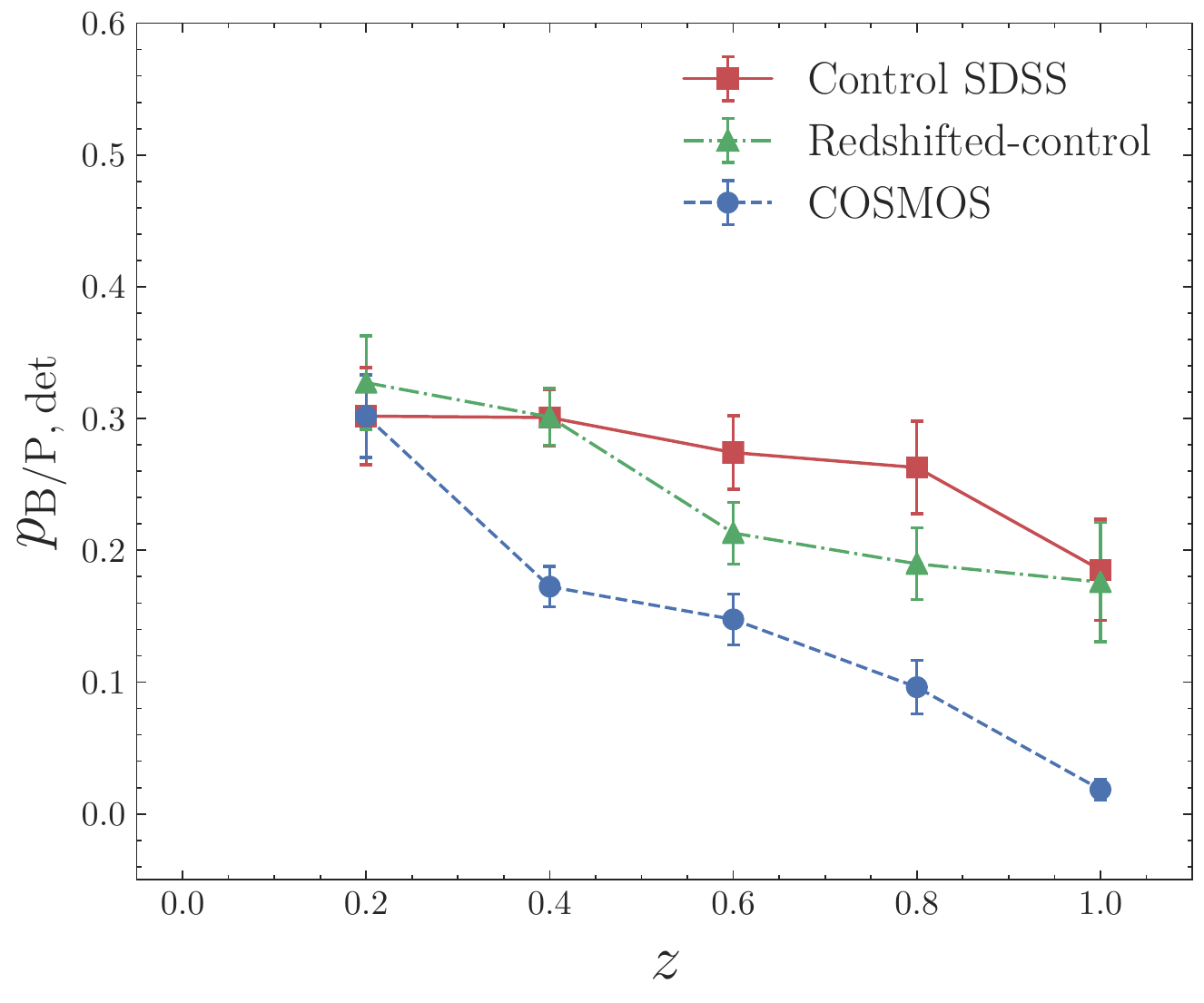}
 \caption{Similar to Figure \ref{BP_vs_redshift}, but with a B/P `likelihood' for each galaxy calculated using equation \ref{BP_likelihood}. The blue points represent the mean B/P bulge likelihood of COSMOS galaxies in each redshift bin. The red square shows the local B/P bulge likelihood as measured for the SDSS galaxies, and the green triangles the comparison, \textsc{redshifted}-control sample to the equivalent redshifts of the COSMOS galaxies. The errors bars represent the standard deviation on the mean, in each bin.}
 \label{BP_probability_vs_redshift}
\end{figure}

\begin{figure}
\centering
 \includegraphics[width=1\columnwidth]{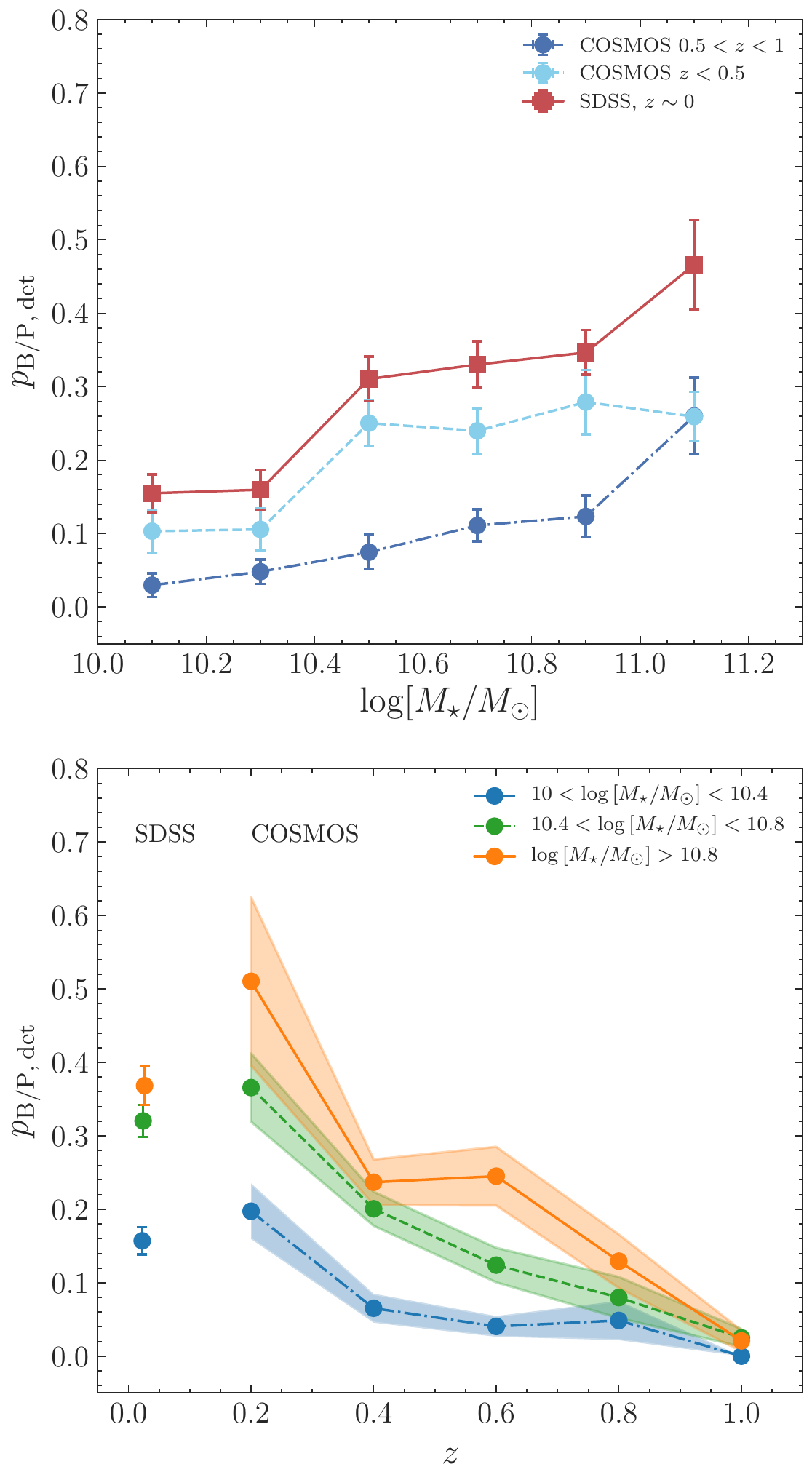}
 \caption{Similar to Figure 
 \ref{HST_SDSS_BP_vs_mass}, but with a B/P `likelihood' calculated using equation \ref{BP_likelihood}. \textit{Top panel}: The blue points represent the mean B/P bulge likelihood of COSMOS galaxies in mass bins, split into two redshift bins ($z<0.5$ and $z>0.5$). The red squares shows the local B/P bulge likelihood in mass bins, as measured for the SDSS galaxies. \textit{Bottom panels}: The redshift evolution of $p_{\mathrm{B/P,det}}$ for three COSMOS mass bins and the value of the local B/P bulge likelihood for the SDSS galaxies, for the same three mass bins. The errors bars represent the standard deviation on the mean, in each bin.}
 \label{BP_probability_mass}
\end{figure}

\noindent{Galaxies with three votes for `Maybe' have, for example, a B/P bulge likelihood of 0.25, while galaxies with three votes for `Probably', 0.75. In Figure \ref{BP_probability_vs_redshift} we plot the average detected B/P likelihood with redshift, for the COSMOS sample, the SDSS sample and the \textsc{redshifted}-control sample. The trend of $p_{\mathrm{B/P,\:det}}$ with redshift is very similar to that of $f_{\mathrm{B/P,\:det}}$, shown in Figure \ref{BP_fraction_comparison}, for all three samples. This is due to the relatively good agreement in the classification of B/P bulges of the three authors. The mean detected B/P likelihood decreases with redshift for the COSMOS sample, from 30\% at $z=0.2$ to 10\% at $z=0.8$ and $\sim3\%$ at $z=1$. The fact that the B/P likelihood is the same for all datasets at $z=0.2$ and that the gradient is steeper for COSMOS than for the comparison \textsc{redshifted}-control dataset, suggests that the observed evolution in the B/P likelihood is genuine, and not due to observational biases. Furthermore, the trends of the mean B/P likelihood with stellar mass and redshift (Figure \ref{BP_probability_mass}) are similar to what we observed before, in Figure \ref{HST_SDSS_BP_vs_mass}, with an increase in $p_{\mathrm{B/P,\:det}}$ at $M_{\star}\gtrsim10^{10.4}\:M_{\odot}$, observable especially in SDSS and in the first COSMOS bin ($z<0.5$). }

Therefore, the results that the fraction of galaxies hosting B/P bulges evolves with redshift and that there is a strong dependence of the presence of a B/P bulge with galaxy stellar mass, at all redshifts, are similar to what we previously found and do not depend on our choice of threshold for classifying galaxies into having thin bars or B/P bulges. 

\section{B/P bulges at $\lowercase{z}>1$}
\label{candels}

We investigated the presence of B/P bulges up to $z=1$. This redshift limit was selected because at higher redshifts the filter used in the COSMOS survey (F814W on the $HST$ ACS Camera) probes the rest frame UV, in which clumpy star formation dominates and bars become invisible in the isophotes because they host mainly older stellar populations. Therefore, there is a possibility that B/P bulges exist at higher redshifts. Indeed, even though we did not detect a B/P bulge in COSMOS at $z=1$, this bin contains few galaxies. The linear fit to the corrected B/P fractions (Figure \ref{f_BP_corrected}) suggests a B/P bulge fraction of $\sim10\%$ at $z=1$ and also the B/P likelihood in Figure \ref{BP_probability_vs_redshift} is non-zero ($\sim3\%$). 

The only existing multiband survey, with sufficient depth and spatial resolution is the $\textit{HST}$ CANDELS survey \citep{Grogin2011,Koekemoer2011}. CANDELS combines optical and near-infrared observations from the ACS and Wide Field Camera 3 (WFC3), respectively, covering an area of $\sim0.2$ deg$^{2}$ from five other surveys (GOODS-North and -South, \citealt{Giavalisco2004}, EGS, \citealt{Davis2007}, UDS, \citealt{Cirasuolo2007} and COSMOS). $\textit{HST}$ CANDELS survey is much smaller in area compared to COSMOS, but $\sim$2 mag deeper. The majority of the fields were imaged with the I filter (F814W), the same as in the COSMOS survey, and, additionally, in the Y (F105W), J (F125W), and H (F160W) filters. Furthermore, the resolution of the WFC3 H-band is twice lower than that of the ACS F814W (0.18$^{\prime\prime}$ compared to 0.09$^{\prime\prime}$). Morphologies for the galaxies are available from the Galaxy Zoo: CANDELS project \citep{Simmons2017}, and photometric redshifts are available from \citet{Dahlen2013}.

\begin{figure}
\centering
 \includegraphics[width=\columnwidth]{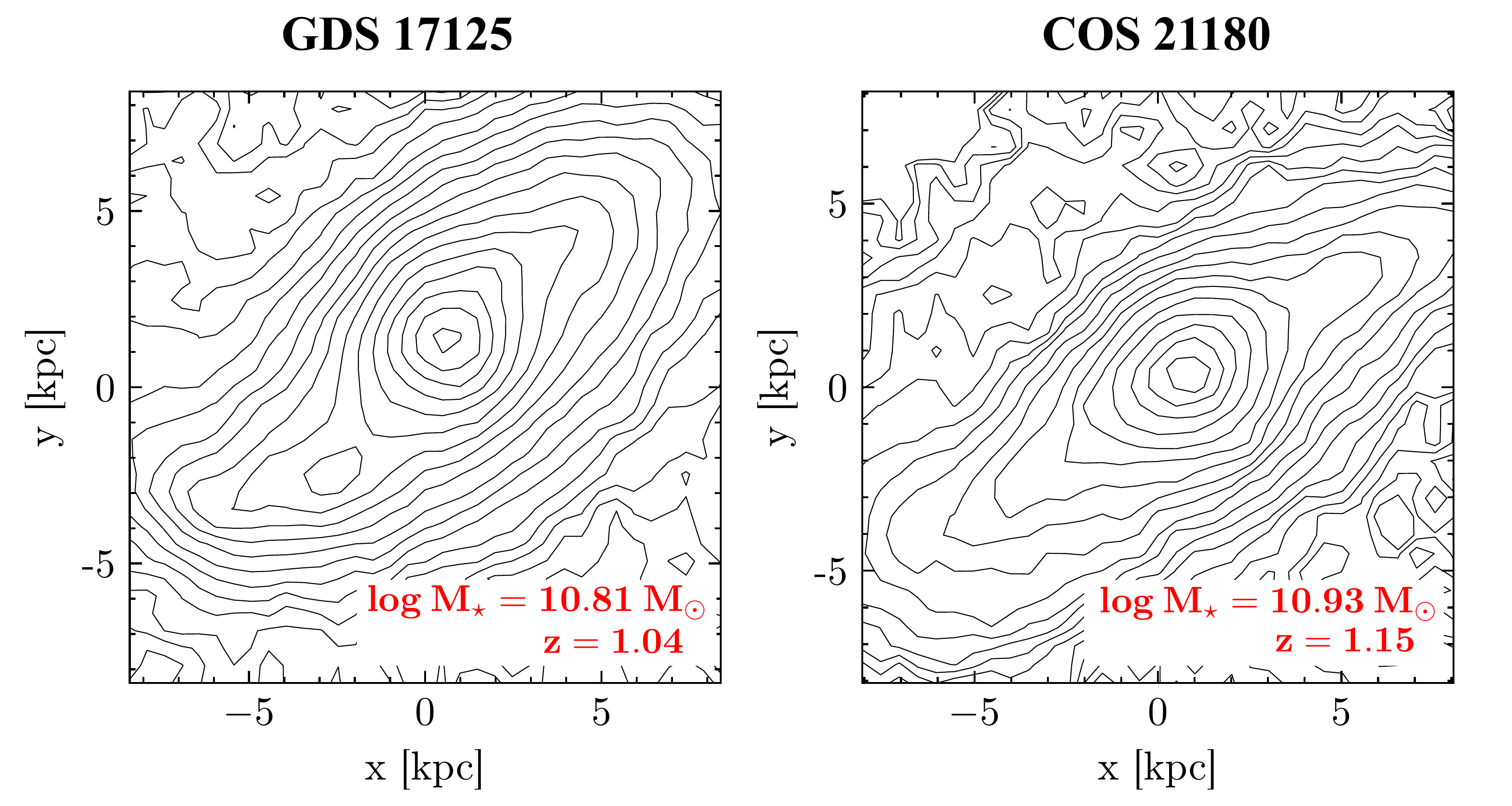}
 \caption{Two galaxies possibly hosting B/P bulges at $z\gtrsim1$, identified in the $HST$ CANDELS survey, in the $H$ band.}
 \label{example_candels_z1}
\end{figure}

Using a similar selection of barred galaxies as for the COSMOS subset (Section \ref{COSMOS_data}), we selected 63 barred galaxies with $M_{\star}>10^{10} M_{\odot}$ and inspected their isophotes in the $H$ band, which probes the rest frame near-infrared in the nearest galaxies, and the rest frame $I$ band for galaxies at $z\gtrsim1$. Inspecting the isophotes of these galaxies, we identified 13 barred CANDELS galaxies possibly hosting B/P bulges, a total fraction of 20.6\%, comparable to the total detected fraction of 17\% found in COSMOS (up to $z=1$). Additionally, we identified two galaxies \textit{possibly} hosting B/P bulges at $z\approx1$ (see Figure \ref{example_candels_z1}), a fraction of 10\% of galaxies at $z\gtrsim1$, in agreement with our previous estimate from the linear fit in Figure \ref{f_BP_corrected}. This suggests that a very small fraction of galaxies might have formed B/P bulges by $z\approx1$, 8 Gyr ago. Considering that the highest redshift barred galaxy was observed at $z\approx2$ (also in CANDELS; \citealt{Simmons2014}), it is plausible that these galaxies formed B/P bulges by $z=1$, within 1-2 Gyr after bar formation, as simulations suggest \citep{Martinez-Valpuesta2006}.

\paragraph*{ACKNOWLEDGEMENTS.} 

The authors would also like to thank the anonymous referee for their helpful and useful comments. We would also like to thank S\'ebastien Comer\'on for the useful discussions on thin and thick discs. This publication has been made possible by the participation of more than 200,000 volunteers in the Galaxy Zoo 2 and Galaxy Zoo: Hubble projects. Their contributions are individually acknowledged at \texttt{http://authors.galaxyzoo.org}. Galaxy Zoo 2 was developed with the help of a grant from The Leverhulme Trust. 

This publication uses data generated via the Zooniverse.org platform, development of which is funded by generous support, including a Global Impact Award from Google, and by a grant from the Alfred P. Sloan Foundation.

VPD was supported by Science and Technology Facilities Council Consolidated grant \#~ST/R000786/1.  The $N$-body simulation in this paper was run at the DiRAC Shared Memory Processing system at the University of Cambridge, operated by the COSMOS Project at the Department of Applied Mathematics and Theoretical Physics on behalf of the STFC DiRAC HPC Facility (www.dirac.ac.uk). This equipment was funded by BIS National E-infrastructure capital grant ST/J005673/1, STFC capital grant ST/H008586/1 and STFC DiRAC Operations grant ST/K00333X/1. DiRAC is part of the National E-Infrastructure.  

Funding for the SDSS and SDSS-II has been provided by the Alfred P. Sloan Foundation, the Participating Institutions, the National Science Foundation, the U.S. Department of Energy, the National Aeronautics and Space Administration, the Japanese Monbukagakusho, the Max Planck Society, and the Higher Education Funding Council for England. The SDSS Web Site is http://www.sdss.org/. 

The SDSS is managed by the Astrophysical Research Consortium for the Participating Institutions. The Participating Institutions are the American Museum of Natural History, Astrophysical  Institute Potsdam, University of Basel, University of Cambridge, 
Case Western Reserve University, University of Chicago, Drexel University, Fermilab, the Institute for Advanced Study, the Japan 
Participation Group, Johns Hopkins University, the Joint Institute for Nuclear Astrophysics, the Kavli Institute for Particle Astrophysics and Cosmology, the Korean Scientist Group, the Chinese Academy of Sciences (LAMOST), Los Alamos National Laboratory, the Max-Planck-Institute for Astronomy (MPIA), the Max-Planck-Institute for Astrophysics (MPA), New Mexico State University, Ohio State University, University of Pittsburgh, University of Portsmouth, Princeton University, the United States Naval Observatory and the University of Washington. 

This research made use of NASA's Astrophysics Data System Bibliographic Services. This work made extensive use of \textit{Astropy}\footnote{\url{http://www.astropy.org/}}, a community-developed core Python package for Astronomy \citep{Astropy} and of the Tool for Operations on Catalogues And Tables, \citep[TOPCAT\footnote{\url{http://www.star.bris.ac.uk/~mbt/}};][]{Topcat}.

\bibliographystyle{mnras}
\bibliography{references}




\bsp	
\label{lastpage}
\end{document}